\documentclass{article}

\usepackage{arxiv}

\usepackage[utf8]{inputenc} 
\usepackage[T1]{fontenc}    
\usepackage{hyperref}       
\usepackage{url}            
\usepackage{booktabs}       
\usepackage{amsfonts}       
\usepackage{nicefrac}       
\usepackage{microtype}      
\usepackage{lipsum}

\usepackage{hyperref}
\usepackage{xcolor}
\usepackage{gensymb}
\usepackage{graphicx}
\usepackage{subfig}
\usepackage{geometry}
\usepackage{comment}
\usepackage{soul}
\usepackage{amsmath}
\usepackage{algorithm}
\usepackage{algorithmic}
\usepackage{mathrsfs}
\usepackage{siunitx}
\usepackage{multicol}
\usepackage{changepage}
\usepackage{ragged2e}

\newlength\savedwidth

\newcommand\thickhline{\noalign{\global\savedwidth\arrayrulewidth\global\arrayrulewidth 2pt}%
\hline
\noalign{\global\arrayrulewidth\savedwidth}}

\title{Suicide disparities across urban and suburban areas in the U.S.: A comparative assessment of socio-environmental factors using a data-driven predictive approach}

\author{
  Sayanti Mukherjee\thanks{Assistant Professor, Director of OASIS Laboratory, Department of Industrial and Systems Engineering, University at Buffalo, The State University of New York} \thanks{Corresponding Author: 411 Bell Hall, Buffalo NY 14260; Email: sayantim@buffalo.edu; Phone: 716- 645-4699}\\
    Department of Industrial and Systems Engineering\\
    University at Buffalo\\
    Buffalo, NY, 14260\\
  \texttt{sayantim@buffalo.edu} \\
   \And
 Zhiyuan Wei \\
  Department of Industrial and Systems Engineering\\
    University at Buffalo\\
    Buffalo, NY, 14260\\
  \texttt{zwei7@buffalo.edu} \\
}

\begin{document}
\maketitle

\begin{abstract}
Disparity in suicide rates between urban and suburban/rural areas is growing, with rural areas typically witnessing higher suicide rates in the U.S. While mental health plays a key role in suicides, other factors such as cultural, socioeconomic, and environmental facets are also influential. However, previous studies often ignored the effect of socio-environmental factors on the suicide rates and its regional disparity. To address these gaps, we propose a holistic data-driven framework to model the associations of social (demographic, socioeconomic) and environmental (climate) factors on suicide rates, and study the disparities across urban and suburban areas. Leveraging the county-level suicide data from 2000---2017 along with the socio-environmental features, we trained, tested and validated a suite of advanced statistical learning algorithms to identify, assess and predict the influence of key socio-environmental factors on suicide rates. We assessed models’ performance based on both in-sample goodness-of-fit and out-of-sample predictive accuracy to ensure a high generalization performance. Random forest outperformed all other models in terms of goodness-of-fit and predictive accuracy, and selected as the final model to make inferences. Our results indicate that population demographics is significantly associated with both urban and suburban suicide rates. We found that suburban population is more vulnerable to suicides compared to urban communities, with suburban suicide rate being particularly sensitive to unemployment rate and median household income. Our analysis revealed that suicide mortality is correlated to climate, showing that urban suicide rate is more sensitive to higher temperatures, seasonal-heating-degree-days and precipitation, while suburban suicide rate is sensitive to only seasonal-cooling-degree-days. This work provides deeper insights on interactions between key socio-environmental factors and suicides across different urbanized areas, and can help the public health agencies develop suicide prevention strategies to reduce the growing risk of suicides.
\end{abstract}


\section{Introduction}
Suicide rates have increased approximately 30\% in the U.S. since 1999 and have become the tenth leading cause of death nationwide, rendering it to be a grievous concern in public health nationally and internationally  \cite{bachmann2018epidemiology, stone2018vital, NationalInstituteofMentalHealthNIMH2019}. Particularly, studies demonstrated that there is a growing disparity in suicide rates between urban and rural areas in the U.S., and highlighted that more urban areas typically witness lower suicide rates while less urban areas experience higher suicide rates \cite{kegler2017trends,hedegaard2018suicide}. Additionally, statistics showed that age-adjusted suicide rate in the remote rural counties of the U.S. was reported to be 1.8 times higher than the populous urban counties (as of 2017) \cite{ivey2017suicide,hedegaard2018suicide}. 
With the overall continued urbanization trend in population migration from rural areas to urban areas of the U.S.\cite{johnson2019rural}, it has become even more critical to understand why the less urban regions are more vulnerable to mental health issues, and identify the key factors that are significantly associated with such higher suicide mortality rates.

Suicidal behavior is considered to be an outcome of the interactions among a number of factors, ranging from personal genomics (a.k.a. internal factor associated with individual characteristics) to environmental influences (a.k.a external factors). 
Some researchers conducted longitudinal study to examine the mechanisms that transmitted the suicidal behaviors from parents to children, in order to discover the presence of heritability of suicidal behaviors through families. The research concluded that adults whose parents had suicidal acts were vulnerable to suicides, with a nearly five times higher likelihood of exhibiting suicidal behaviors compared to an average person \cite{brent2002familial,brent2015familial}. 
In additional to genetics, environmental factors and traumatic events are also associated with higher rates of suicide and suicidal thoughts. Suicide mortality rates are influenced by multifaceted environmental factors such as social, economic and demographic characteristics of a population  \cite{huang2017demographics}. Recently, with the growing concerns of global warming, some researchers examined the relationship between climate conditions and suicidal behaviors, but the conclusions were contradictory. Some of studies indicated that a higher suicide rate is positively correlated with the elevated temperature \cite{dixon_association_2014, burke2018higher}, on the contrary, others found that an increase in suicide rate is linked to a lower temperature \cite{preti1998influence, lester1999climatic, williams2015will}. Thus, it is of particular importance to incorporate climate conditions when examining the suicide trend of the population. 

By leveraging a multifaceted socio-environmental data (e.g., socioeconomic, demographics and climate) collected from publicly available data sources, we propose to develop a holistic data-driven predictive framework leveraging statistical learning theory to model the interactions between various socio-environmental factors and suicide mortality rates across the different urban and suburban areas in the U.S. We aim to identify the key influencing factors that could explain the increasing disparity in suicide rates across the various urbanized regions, and evaluate their associations with the differential suicide trends in the urban and suburban regions.

The notable contributions of our study is threefold. First, a wider range of variables defining socio-environmental conditions including socio-economic condition of the population, demographics and climatic factors has been examined in relation to suicide rates across the urban and suburban counties in the U.S. using a systematic holistic approach. Second, for the first time, a robust data-driven framework leveraging a set of statistical models is proposed to model the complex associations between the socio-environmental factors and the suicide mortality rates. Finally, a comparative assessment of the key factors is provided to evaluate the suicide disparities in the urban (large central metropolitan counties) and suburban (medium/small metropolitan counties) counties.

\section{Background}
Suicide is complex, multi-factorial behavioral phenotype. It is considered to be an outcome of complex interactions between a multitude of internal (e.g., personal characteristics, mental and physical illness) and external entrapment (e.g., environmental factors,  traumatic events) \cite{lewitzka2016personality}. A large body of literature has been investigated the relationship between internal factors and suicidal behaviors at the individual-level. However, since the purpose of this study is, at the population level, to understand and evaluate the environmental effects on suicide mortality rates across urban and suburban counties, the literature review presented in this section mostly focus on socioeconomic, demographic and climatic factors related to suicides and suicidal behaviors. 

\subsection{The socioeconomic and demographic factors}
In the literature, some researches examined the difference in suicide rates between males and females, and observed the well-known philosophy of ``gender paradox'' in suicide---i.e., females typically have higher rates of suicide ideation, but lower rates of suicide mortality compared to males \cite{canetto1998gender,moore2018gender}. In addition to gender, other demographic factors can also play a critical role in linking to a higher suicide risk.
A meta-analysis was performed to highlight that although demographic factors were found to be statistically significant, they were weak (i.e., no single demographic factor appeared to be particularly strong) in contributing to the overall complex phenomenon of suicidal behaviors \cite{huang2017demographics}. Additionally, the authors suggested that further studies are needed to understand the effects of demographics on suicide mortality rates. Another nationwide study conducted for Iran from 2006 to 2010 concluded that certain demographic factors such as gender, age and education level could influence people in adopting different methods to commit the suicide \cite{shojaei_association_2014}. The authors found that younger generation was more likely to use a highly violent method such as firearms to complete suicide, while the elderly people often selected hanging and poisoning as means to commit suicide \cite{shojaei_association_2014}. In addition, it was found that men preferred hanging while women preferred self-burning to end their lives \cite{shojaei_association_2014}. The authors also concluded that hanging was more prevalent among low educated people while poisoning was more popular among higher educated groups \cite{shojaei_association_2014}. 

The suicide mortality rate varies significantly among different racial and ethnic groups. A previous study concluded that African Americans were more likely to select violent methods in committing suicides than Caucasians, when socioeconomic status and other factors were kept constant \cite{stack2005race}. 
Other studies revealed that the adolescent minority groups such as Native Hawaiian/Pacific Islander and American Indian/Alaska Native as well as multi-racial groups were highly vulnerable to committing suicides, compared to their Asian, Black, Hispanic, and White counterparts \cite{wong2012ethnic,wei2020health}.
The educational attainment can be also linked to suicide risk. For example, people with a college degree or higher, exhibited lowest rates of suicide, whereas those with a high school diploma only were found to be more vulnerable with an increased risk of suicide \cite{phillips2017differences}. Similarly, the results from another study pointed out that men with lower educational attainment had a higher risk of suicide in eight out of ten European countries, while suicide rates among women was found to be low and less consistent across all the countries \cite{pompili_does_2013}.   

Economic condition also plays a critical role in affecting suicide mortality rates. This is expected and it is established that poor socio-economic conditions characterized by higher incidence of poverty, lack of health insurance and higher unemployment rates are critical in affecting the mental health and wellbeing of adults in metropolitan areas \cite{mukherjeetowards}. For instance, a study conducted in Taiwan to explore the relationship between unemployment rate and suicide rate, found that a 1\% increase in absolute unemployment rate was linked to a 4.9\% increase in the relative age-adjusted suicide rate from 1978-2006 \cite{chen_suicide_2010}. Suicide rate was found to be statistically different across genders---men were found to be more likely to commit suicides compared to women in face of economic turmoil and financial issues \cite{vandoros2019association}. This gender difference in suicide mortality rate during economic crisis is also unwrapped in another study, where researchers found that men with lower per capita income more frequently committed suicides, while such a phenomenon was not observed in the female group \cite{sher2006per}. 

Previous exploratory data analysis on suicide rates in rural and urban counties in the U.S. revealed that the age-adjusted suicide rate for most of the rural counties was 1.8 times higher than most of the urban counties in 2017, and its rate had been rapidly increasing over the past decade. The authors, however, neither attribute a cause to such an increase in the suicide rates, nor did it explain why the difference was observed between rural and urban counties \cite{hedegaard2018suicide}.

\subsection{The climatic factors}
To understand the impact of environmental factors on suicide rates, some researchers investigated the weather-induced higher risk of suicide. A systemic literature review was conducted to highlight that air temperature had a significant influence on suicidal acts, but their correlation could be either positive or negative, depending on the variation of sociological or geographic factors across different populations \cite{deisenhammer2003weather}.  
In another study \cite{dixon_association_2014}, the authors established a distributed lag nonlinear model (DLNM) to determine the relationship between suicide rates and air temperature in Toronto, Ontario (Canada), and Jackson, Mississippi (USA). The models from both the locations concluded that warmer than normal temperatures had a positive correlation with the total number of suicides. However, the authors claimed that since the data was only from two cities, it might not be sufficient to establish immediate clinical implications, but can guide further investigations to better understand and quantify the suicide rates associated with temperature changes \cite{dixon_association_2014}.
A positive correlation between elevated temperatures and suicide rates had also been established in another nationwide study \cite{burke2018higher}, where the authors analyzed decades of historical data (1968–2004 in the U.S. and 1990-2010 in Mexico) and demonstrated that the relationship between temperature and suicide was roughly linear using distributed lag models. It was observed that a 1$^{\circ}$C increase in average monthly temperature could contribute to an increase in the monthly suicide rate by 0.68\% in the U.S., and 2.1\% in Mexico. The study projected that under climate change, suicide mortality rate would increase by 1.4\% in the U.S. and 2.3\% in Mexico by 2050 \cite{burke2018higher}.
On the contrary, a negative correlation between temperature and suicide rates was observed in the other studies \cite{preti1998influence, lester1999climatic, williams2015will}, suggesting that decreasing temperature was linked to a growing rate of suicide incidents. The cause of this contradiction might be explained as climate variation could have heterogeneous effects across geographic areas \cite{helama2013temperature}. In this view, the further studies are needed to investigate the complex interactions of climate-induced shifts in suicidal behaviors by controlling other factors such as spatiotemporal and socioeconomic backgrounds.

\subsection{Existing research gaps and our contribution}
Previous research studies have been investigated the impacts of certain environmental factors in relation to suicide risks. However, some knowledge gaps still exist that are summarized below. 

\begin{enumerate}
    \item Most of the existing studies independently examined the relationships of socioeconomic, demographic and climate factors in relation to the growing risk of suicide, with lack of consideration on the interdependence of these factors on suicide rates. Suicide is a complex phenomenon that cannot be adequately captured by a single feature or feature type. 
    \item Disparities in suicide mortality rate across different types of urbanized areas are paid less attention. Although previous studies indicated that the difference in suicide rates between urban and rural areas in the country has been widening, with higher rates observed in less urban areas, they did not provide insights on potential factors that can explain such a difference \cite{hedegaard2018suicide}. 
    \item Most of the previous studies applied the traditional linear models and basic statistical analyses (e.g., Pearson correlation coefficient) to characterize the relationship between the potential risk factors and the increased risk of suicide. These traditional approaches, however, fail to adequately capture the nonlinear characteristics in the complex structure of data in modeling suicide risks.
    \item Moreover, in the face of climate change and growing urbanization, the models' strong capability of predicting suicide risks is particularly critical. The robust predictive approach in modeling suicide risks has been attracting little attention in the previous studies.
\end{enumerate}

In order to address the above-mentioned existing gaps, our study examined the impact of socio-environmental factors (i.e., integration of socioeconomic, demographics and climate features) on suicide disparities across the large central and medium/small metropolitan areas in the U.S. during the period of 2000---2017, and proposed a holistic data-driven predictive approach to model the relationship between socio-environmental factors and suicide rates, leveraging a library of advanced statistical learning techniques. Finally, a comparative assessment of the key influencing factors is implemented to evaluate the suicide disparities across different urbanized regions.

\section{Data collection, preprocessing and visualization} \label{section:data_collection}
In this section, we present the data collected from  multiple publicly available sources, and a sequence of data preprocessing steps to clean and aggregate the data, as well as data visualization to provide a basic understanding of suicide rates across different spatial-temporal scales. 

\subsection{Data collection}
Suicide mortality data was collected from the Centers for Disease Control and Prevention (CDC) for the period 2000---2017 monthly using CDC's WONDER tool \cite{CDCwonder} based on the variables---\textit{County, Month, Year, Intent of injury}. ``Intent of injury" describes an act of injury caused on purpose by oneself or by another person, with the goal of injuring or killing themselves or others \cite{CDCinjury}. Here, the intent of injury is specified as ``suicide'' in this study. 

The urbanization level of a county was determined using the six-level urban-rural classification provided by the latest 2013 National Center for Health Statistics (NCHS) Urban–Rural Classification Scheme for Counties \cite{ingram20142013, hedegaard2018suicide}. This classification scheme distinguishes six urbanization levels based on metropolitan/non-metropolitan status, population distribution, and other factors. Our study utilized different levels of the urban hierarchy to better explain the suicide disparities between less urban and more urban regions.

In addition, socio-environmental data were gathered as well. The county-level socioeconomic and demographic information were collected from the U.S. Department of Agriculture (USDA) Economic Research Service (ERS) \cite{USDA_ERS} for the period of analysis from 2000---2017. And, climate data was obtained from the National Oceanic and Atmospheric Administration's (NOAA) National Climatic Data Center (NCDC) \cite{NOAA}. The climate data captures several weather variations on a monthly basis from 2000---2017. 

\subsection{Data preprocessing and aggregation}
\begin{figure}[!h]
\centering
\includegraphics[width=0.7\textwidth]{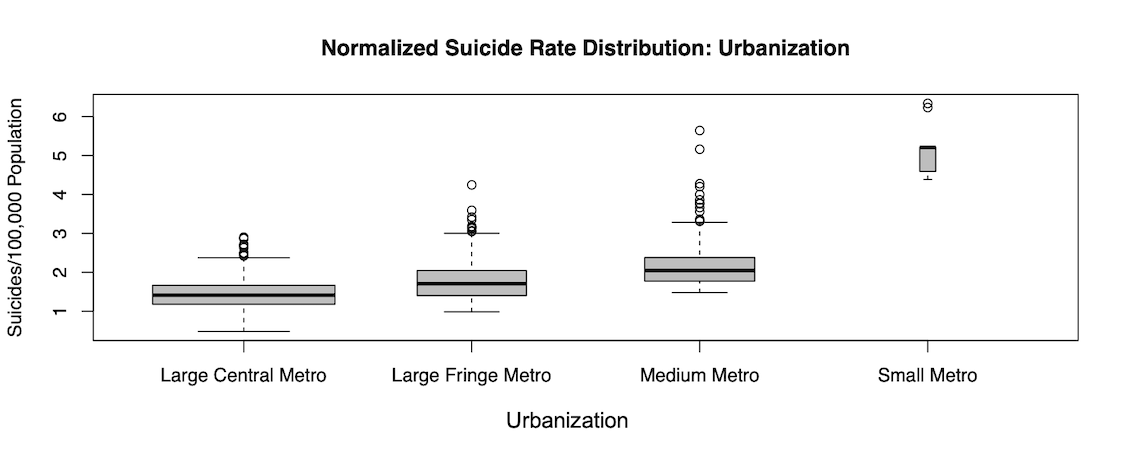}
\caption{{\bf Box plot showing normalized suicide mortality rates across various metropolitan counties.}
Metropolitan counties classifications area based on NCHS Urban–Rural Classification Scheme for Counties \cite{ingram20142013}. The width of each of the box plots indicates the number of observations in each categories.}
\label{fig2}
\end{figure}

Data directly collected from multiple sources contain some missing or wrong entries. Specifically, CDC WONDER disabled public reporting of the suicide incidents in a county where the monthly incident count was below the pre-determined ``cut-off'' value \cite{WONDER}, in order to protect confidential information such as event details related to an individual victim. In this view, counties were selected when satisfying: 1) having at least a total of ten reported suicide cases in every month during 2000---2017; and 2) vulnerable to witnessing higher suicide trend consistently. Those selected counties are from the states---Arizona, Colorado, Idaho, Kansas, Kentucky, Missouri, Nevada, New Hampshire, New Mexico, Oklahoma, Oregon, Tennessee, Utah, and Washington. Based on selected metropolitan counties, the bar plot in Fig \ref{fig2} shows the distribution of normalized suicide mortality (per 100,000 of the population) in the four types of metropolitan counties.
In order to balance the number of the counties with different urbanization levels, we combined the large fringe metros, medium metros and small metros into a single category---``medium/small metros'', while keeping the most urbanized areas as ``large central metros''. Moreover, the socioeconomic and demographic characteristics of the large central metros are significantly different than that of the medium/small metros (i.e., large fringe metros, medium metros and the small metros) \cite{parker2018demographic}, which also justifies our grouping criteria. The relevant counties used in our analysis along with their urbanization classification are listed in the Appendix.

County-level data on suicide mortality rates, socioeconomic and demographic information, and climate details were integrated using \textit{year, month} and \textit{county} as the ``common keys'' to connect across different datasets. For socio-environmental factors, the variables are removed when satisfying: 1) more than 20\% of missing inputs; or 2) highly correlated with other variables based on Pearson correlation coefficient ($\rho\geq$0.9 or $\rho\leq$-0.9). Removing those highly correlated variables can avoid the ``masking effect'' of certain variables or model overfitting, and will help with key influencing factors assessment in model inferencing \cite{mukherjee2017climate, obringer2020evaluating}. Finally, the socio-environmental variables used in this study include three socioeconomic variables, seventeen demographic variables, eight climatic variables and one binary variable (urbanization level indicates either the target is a large central or medium/small metropolitan area). Those variables are explained in Table \ref{table:socioenvironmental}. Additionally, three spatiotemporal variables used to link the different datasets are explained in Table \ref{table:keyVars}.

After variable pre-selection and data aggregation, we normalized the suicide mortality counts to suicides mortality rates per 100,000 population, to eliminate the effect of population size in a county. Finally, the final dataset included 2,496 observations and 33 variables including the normalized suicide rates as response variable.

\begin{table}[!ht]
\begin{adjustwidth}{0in}{0in} 
	\begin{center}
	\caption{\bf{Description of socio-environmental variables.}}
	\label{table:socioenvironmental}
	\begin{tabular}{l p{10.8cm} l}
		\hline
		Variable Name & Description & Periodicity \\
		\thickhline
		Urbanization level & Large central metro or medium/small metro per county. & Annually \\
		\hline
		Unemployment Rate & Percent of unemployed workers in the total labor force. & Monthly\\ 
        Poverty & Percent of people (of all ages) in poverty in the county. & Annually \\
        Income & Median household income in the county. & Annually \\ 
        \hline
        Age Group 1 & Percent of county's population ages below 14. & Annually \\ 
        Age Group 2 & Percent of county's population between ages 15--29. & Annually \\ 
        Age Group 3 & Percent of county's population between ages 30--44. & Annually \\ 
        Age Group 4 & Percent of county's population between ages 45--59. & Annually \\ 
        Age Group 5 & Percent of county's population between ages 60--74. & Annually \\ 
        Age Group 6 & Percent of county's population ages above 75. & Annually \\ 
        Female & Percent of county's population female. & Annually \\ 
        NA & Percent of county's population which is Native Hawaiian, Pacific Islander alone (i.e., no other race). & Annually \\ 
        AA & Percent of county's population which is Asian alone. & Annually \\ 
        IA & Percent of county's population which is American Indian, Alaska native alone. & Annually \\
        BA & Percent of county's population which is Black alone. & Annually \\
        WA & Percent of county's population which is White alone. & Annually \\
        NH & Percent of county's population which is non-Hispanic. & Annually \\
        Education Group 1 & Percent of county's population whose education level is less than a High School diploma. & Annually \\
        Education Group 2 & Percent of county's population whose education level is a High School diploma only. & Annually \\
        Education Group 3 & Percent of county's population whose education level is some college of Associates degree. & Annually \\
        Education Group 4 & Percent of county's population whose education level is a Bachelor's degree or higher. & Annually \\
        \hline
        DP10 & Number of days with $\ge$ 1.00 inch of precipitation in the month. & Monthly \\
		DT00 & Number of days with minimum temperature $\le$ 0 degrees Fahrenheit. & Monthly \\
		DX32 & Number of days with maximum temperature $\le$ 32 degrees Fahrenheit. & Monthly \\
		DX70 & Number of days with maximum temperature $\ge$ 70 degrees Fahrenheit. & Monthly\\
		DX90 & Number of days with maximum temperature $\ge$ 90 degrees Fahrenheit. & Monthly\\
		EMXP & Extreme maximum daily precipitation total within month. Values are given in inches (to hundredths). & Monthly \\
		CDSD & Cooling degree days (season-to-date). Running total of monthly cooling degree days through the end of the most recent month. Each month is summed to produce a season-to-date total. Season starts in July in Northern Hemisphere and January in Southern Hemisphere. & Monthly\\
		HDSD & Heating degree days (season-to-date). Running total of monthly heating degree days through the end of the most recent month. Each month is summed to produce a season-to-date total. Season starts in July in Northern Hemisphere and January in Southern Hemisphere. & Monthly\\ 	\hline
	\end{tabular}
	\end{center}
	\end{adjustwidth}
\end{table}

\begin{table}[!ht] 
\begin{adjustwidth}{0in}{0in} 
	\begin{center}
	\caption{\bf{Description of key variables used to connect all datasets}}
	\label{table:keyVars}
	\begin{tabular}{l p{10.5cm}}
		\hline
		Variable Name & Description \\
		\thickhline
		County & County of reported suicide count. \\
		Year & Year of the reported suicide count.\\
		Month & Month of the reported suicide count.\\ \hline
	\end{tabular}
	\end{center}
	\end{adjustwidth}
\end{table}

\subsection{Data visualization}
\begin{figure}[!h]
\centering
\includegraphics[width=0.5\textwidth]{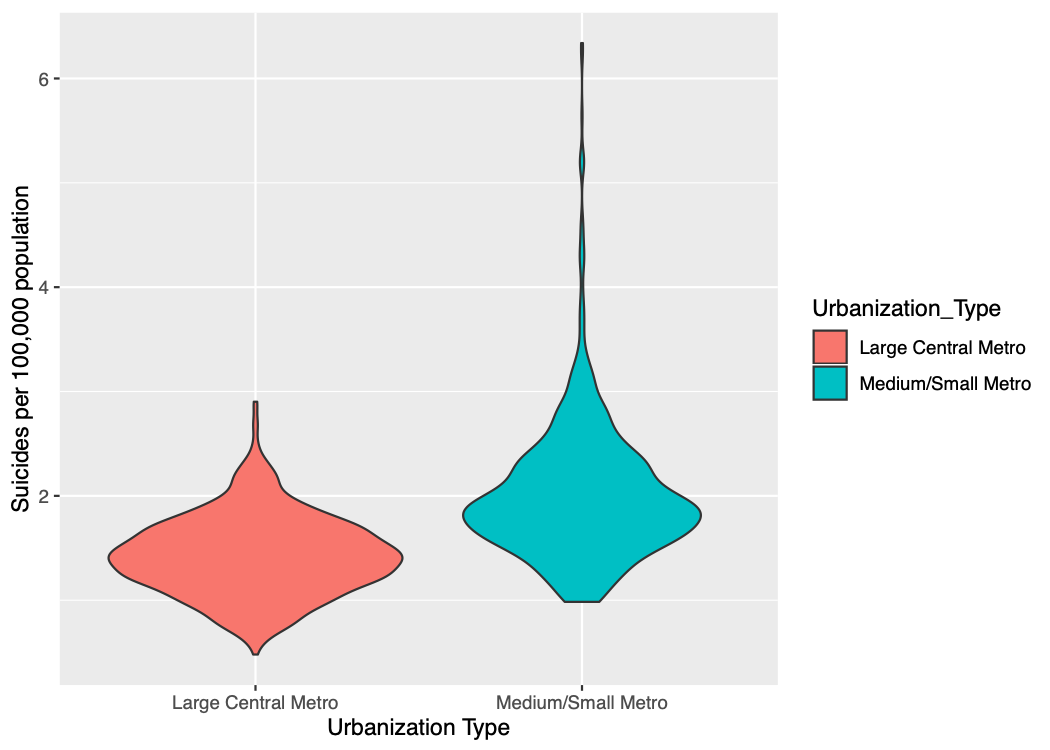}
\caption{{\bf Violin plot depicting normalized suicide mortality rates across large central vs. medium/small metropolitan areas.}
Violin plots are similar to box plots, with a rotated kernel density plot on each side showing the probability density of the data at different values.}
\label{fig3}
\end{figure}

Fig \ref{fig3} exhibits a violin plot depicting the distribution of the normalized suicide mortality rates between large central metro and medium/small metros. Clearly, the distribution of suicide mortality rate is highly right skewed depicting higher suicide rates in the medium/small metropolitan counties, which can reflect the existence of suicide disparity in level of urbanization.

\begin{figure}[!h]
	\begin{center}
	\includegraphics[width=1\textwidth]{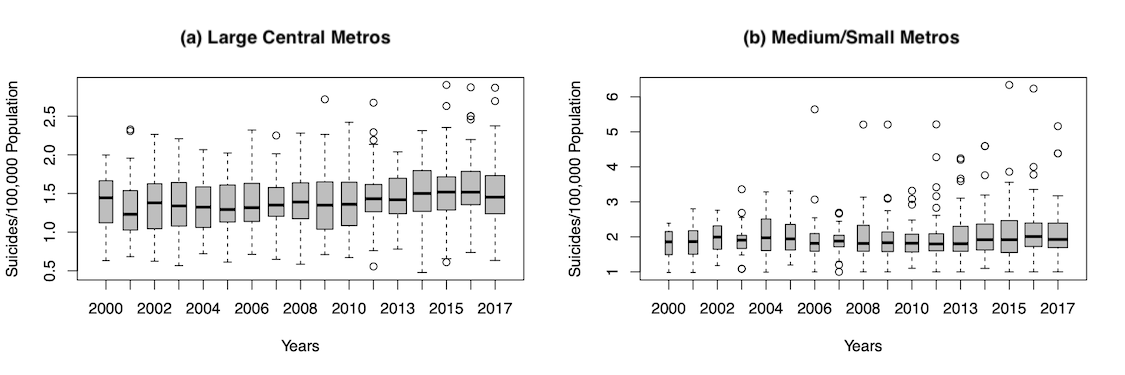}
	\caption{{\bf Box plot showing annual trends in suicide mortality.} (A) Annual distribution of suicide counts per 100,000 population in large central metro; (B) Annual distribution of suicide counts per 100,000 population in medium/small metro areas.}
	\label{fig4}
	\end{center}
\end{figure}

\begin{figure}[!h]
	\begin{center}
	\includegraphics[width=1\textwidth]{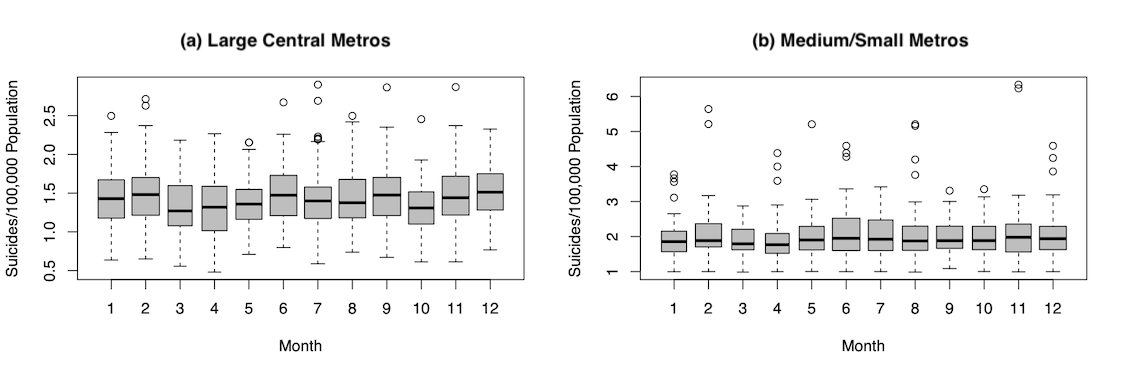}
	\caption{{\bf Box plot showing monthly distribution of suicides.} (A) Monthly distribution of suicides counts per 100,000 population in large central metro; (B) Monthly distribution of suicides counts per 100,000 population in medium/small metro areas.}
	\label{fig5}
	\end{center}
\end{figure}

Fig \ref{fig4} and Fig \ref{fig5} respectively depict the annual trends and monthly distribution in suicide mortality rates in the selected large central and medium/small metropolitan counties in the U.S. As shown from Fig \ref{fig4}(A) and Fig \ref{fig4}(B), there are no significant increasing or decreasing trends in the annual or monthly suicide mortality rates in either of the large and medium/small metropolitan counties. However, more outliers are observed in the post 2000s, indicating suicide rates are increasing in some of the counties. 

\section{Research Methodology} \label{sec:methods}
A data-driven holistic framework for modeling the complex interactions between a number of socio-environmental factors and the growing suicide rates in the large central and medium/small metropolitan areas is explained in this section. Then, we present a brief description of the supervised learning theory where the related predictive model and the statistical techniques used to select the model are also introduced. 

\subsection{Research framework} \label{sec:res_frame}
\begin{figure}[!h]
	\begin{center}
	\includegraphics[width=1\textwidth]{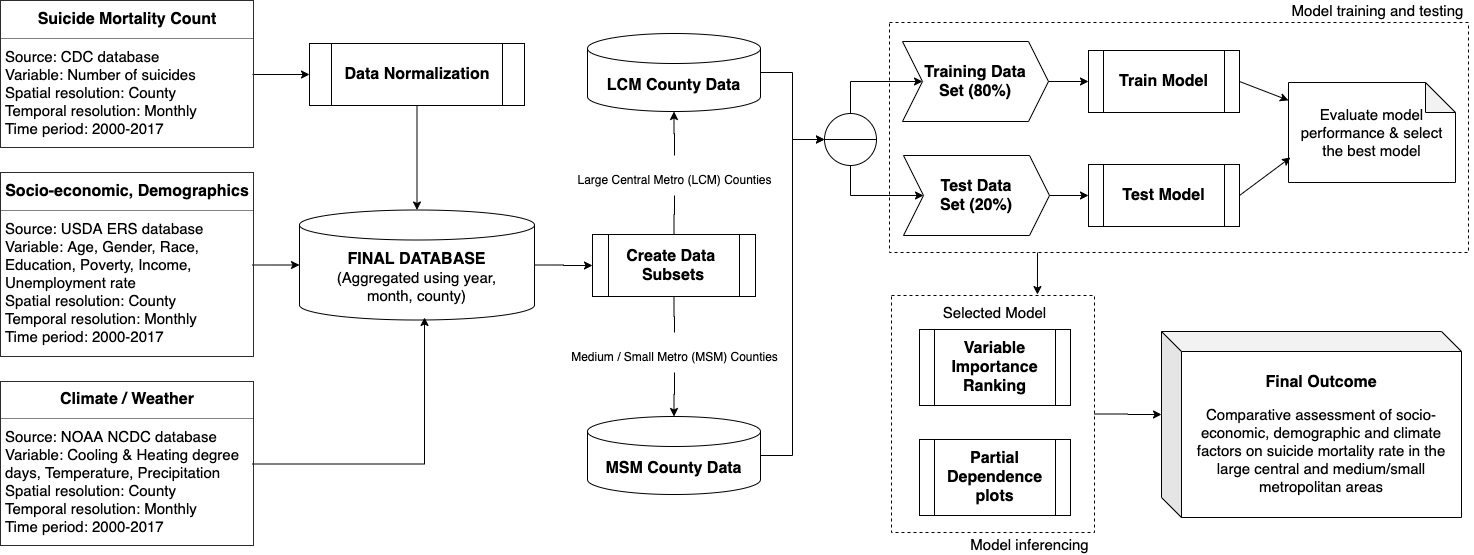}
	\caption{{\bf Schematic of the proposed data-driven research framework.} }
	\label{framework}
	\end{center}
\end{figure}

The schematic of our proposed research framework is exhibited in Fig \ref{framework}. The research framework consists of three major steps: (i) data processing; (ii) model training and testing; and (iii) model inferencing. In Step (i), county-level suicide mortality information and multifaceted socio-environmental variables at different spatiotemporal scales were processed by a series of procedures ranging from data collection, cleaning, normalization, aggregation and visualization. Final aggregated dataset was divided into two independent subsets based on urbanization level---(1) large central metro (LCM), and (2) medium/small metro (MSM). More details in Step (i) can be found in the section of \textbf{Data collection, preprocessing and visualization}. 
Then in Step (ii), a library of regression models were trained and tested separately on each of the LCM and MSM datasets. More specifically, we performed the model training and testing by leveraging a 30-fold 80-20 randomized holdout technique. This technique can be described as follows: in a dataset, 20\% of the data is randomly held-out as test set to evaluate the model's out-of-sample predictive accuracy, while the remaining 80\% of the data is used as training set for training the models. This process is repeated 30 times to ensure that all the data is used at least once to produce generalized errors in training and testing the models \cite{james2013introduction}. The average model performance across all iterations is then calculated in terms of three commonly-used statistical metrics---$R^2$, RMSE (root mean square error) and MAE (mean absolute error). Finally, the model that outperforms other models in terms of out-of-sample predictive accuracy as well as a comparative better goodness of fit is selected as the final model in this paper. The details of final regression algorithm is also elaborated on this section. 
Finally in Step (iii), leveraging the selected model, we analyzed the relative influence of the socio-environmental factors on the suicide mortality rate using the variable importance ranking and partial dependence plots in the setting of both LCM and MSM counties.

\subsection{Supervised learning} \label{sec:supervised}
Supervised learning method is applied to estimate a regression function capable of predicting the response variable $Y$ conditioned on a set of inputs $X$, such that the loss function for measuring errors is minimized. 
The generalized form can be mathematically written as $Y=f(X)+\epsilon$, where $\epsilon$ is the irreducible error follows $\epsilon\sim \mathcal{N}(0,\sigma ^{2})$ \cite{hastie2009elements, james2013introduction}. The loss function $\mathcal{L}$, representing the deviation of observed values from the estimated values of $Y$, typically can be calculated through the absolute error ($L1$ norm) or squared error ($L2$ norm). That is,
$$ 
\mathcal{L}(Y,\widehat{f}(X))=\left\{
\begin{aligned}
\frac{1}{N}\sum_{i=1}^{N} |y_i - \widehat{f}(x_i)| & \qquad \text{mean absolute error (MAE)} \\
\frac{1}{N}\sum_{i=1}^{N} (y_i - \widehat{f}(x_i))^2 & \qquad \text{root mean squared error (RMSE)}, \\
\end{aligned}
\right.
$$
where $N$ is the total number of data points. 

Note that supervised statistical learning models can be parametric, semi-parametric or non-parametric. Parametric models generally assume a particular functional form that relates the input variables to the response. The assumed functional form helps with the ease of estimation and model interpretability, but comes at the cost of  predictive accuracy since the assumptions (such as normality and linearity) often do not hold for real cases. On the other hand, non-parametric models that make no assumption about the distribution of the response variable or the shape of the function relating the response to the predictors, are free to learn any functional form of the response from the training data. By utilizing data in novel ways to estimate the dependencies, the non-parametric models often have a superior predictive power than parametric models. However, the non-parametric methods are data-intensive and highly dependent on the quality of the data. 

In this study, the response variable $Y$ is represented by the normalized suicide rates, and rest of the variables in the dataset are denoted as the predictor variables $X$. The function $f$ is construct through a library predictive models including generalized linear models (GLM) \cite{nelder1972generalized}, ridge and lasso regression \cite{tibshirani1996regression}, generalized additive models (GAM) \cite{hastie1990generalized}, multi-adaptive regression splines (MARS) \cite{friedman1991multivariate}, and ensemble tree based models including random forest (RF) \cite{breiman2001random} and Bayesian additive regression trees (BART) \cite{chipman2010bart}. By implementing a series of experiments, RF outperforms all the models in terms of both goodness of fit and predictive accuracy and thus, we select RF as our final model to assess socio-environmental affects on suicide mortality rates in the metropolitan counties. Details of random forest algorithm and model selection techniques are provided in the following subsections.

\subsubsection{Random forest: algorithm description} \label{sec:RF}
Random forest technique uses a bootstrap aggregating approach combined with feature randomness while building each tree, and attempts to create a multitude of decision trees. For regression problems, the overall model performance is given by averaging predictions from each of the single tree that usually produces low bias yet high variance, to render a more accurate and robust prediction (associated with low bias and low variance). Random forest is an ensemble tree-based learning model that consists of $B$ bootstrapped regression trees $T_{b}$ and explained in details in the Algorithm \ref{algo:RF} \cite{breiman2001random}.
\begin{algorithm}
\caption{Random Forest Algorithm \cite{breiman2001random, james2013introduction}}
\begin{algorithmic}[1]
\label{algo:RF}
\STATE {\textbf{Input:} Data set with dimension ($N,M$) where $N$ is the number of data points \& $M$ is the number of input variables; Ensemble tree size $B$}
\FOR {$b = 1$ to $B$:}  
\STATE Build a bootstrap sample $N_{b}$ from data set of size $N$ by randomly sampling $|N_{b}|$ data points with replacement.
\STATE Treat $N_{b}$ as the training data set, while the remaining data is used as validation set to estimate tree's prediction error.
\STATE Fit a regression tree model $T_{b}$ on the training data set $N_{b}$ by recursively repeating the following steps for each terminal node of the tree, until the minimum node size $n_{\text{min}}$ is reached.
\\ \qquad i) Select $m$ variables randomly from the M variables ($m\ll M$).
\\ \qquad ii) Pick the best variable/split-point among the $m$.
\\ \qquad iii) Split the node into two daughter nodes.
\ENDFOR
\RETURN{} {\{$T_{b}\mid 1\leq b\leq B$\}}
\STATE {\textbf{Output:} Ensemble tree model whose prediction is given by average of predictions across all trees:}
\begin{eqnarray}
\label{eq:rf_model}
	\widehat{f}_{RF}=\frac{1}{B}\sum_{b=1}^{B}T_{b}
\end{eqnarray}
\end{algorithmic}
\end{algorithm} 

\subsubsection{Predictive accuracy vs. model interpretability} \label{sec:modelinter}
Generally speaking, the flexible non-parametric methods have higher predictive power than the ``rigid" parametric methods. However, the improved predictive power comes at the cost of easier interpretability. To make inferences based on non-parametric ensemble tree-based methods, ``partial dependence plots'' (PDPs) are applied to help in understanding the effects of the predictor variable of interest $x_j$ on the response in a ``ceteris paribus” condition to control all the other predictors. Mathematically, the estimated partial dependence can be represented as \cite{chipman2010bart,nateghi2017multi}: 
\begin{eqnarray}
\label{eq:PD}
	\widehat{f_{j}}(x_{j})=\frac{1}{n}\sum_{i=1}^{n}\widehat{f_{j}}(x_{j},x_{-j,i}).
\end{eqnarray}
Here, $\widehat{f}$ represents the statistical model (in this case random forest); $x_{-j}$ denotes all the variables except $x_{j}$; $n$ denotes the number of observations in the training data set. The estimated PDP of the predictor $x_{j}$ provides the average value of the function $\widehat{f}$ when $x_{j}$ is fixed and $x_{-j}$ varies over its marginal distribution.

\subsubsection{Bias variance trade-off and model selection} \label{sec:bias_variance}
Bias variance trade-off is the key to model selection in supervised learning theory. Optimal generalization performance of a predictive model hinges on the ability to simultaneously minimize the bias and variance of the model, thus controlling the complexity of the model. Cross validation is the most widely used technique for balancing models’ bias and variance \cite{james2013introduction}. Thus, we leveraged a percentage randomized holdout technique to estimate the predictive accuracy of the models. More specifically, out-of-sample predictive accuracy of each model was calculated by implementing 30 iterations where in each iteration, 20\% of the data was randomly held out to test model and the model was trained on the remaining 80\% data. The optimal model can be selected in such a way that it outperforms all the other models in terms of in-sample goodness-of-fit and out-of-sample predictive accuracy.

\section{Results} \label{section:results}
In this section, we present a comparative assessment of the in-sample and out-of-sample performances of all the statistical learning models, identify and evaluate the key influencing socio-environmental predictors associated with the suicide mortality rate based on the final model, and compare those factors in contribution to disparity of suicide rate in both the large central and medium/small metropolitan counties in the U.S.
 
\subsection{Comparative assessment of model performance and final model selection} \label{section:model_selection_results}
A summary of the models' performances, developed for both the large central metropolitan and the medium/small metropolitan counties, are provided in Tables \ref{table:LCM_model_perf} and \ref{table:MSM_model_perf} respectively. Performance of the models, in terms of in-sample model fit and out-of-sample predictive accuracy, is evaluated using three statistical metrics (i.e. $R^2$, RMSE, MAE) that averaged across the 30 iterations. Model's in-sample fit indicates it's ability to capture the underlying structure of the data and explains response as a function of the predictors, while the predictive accuracy measures the model's ability to make future predictions. 

\begin{table}[!ht]
\begin{adjustwidth}{0in}{0in} 
	\begin{center}
	\caption{\bf{Large Central Metropolitan Counties: Model performance comparison.}}
	\label{table:LCM_model_perf}
	\begin{tabular}{lp{7.2cm}|ccc|ccc}
	\hline
		\multicolumn{8}{c}{Large Central Metropolitan County Model} \\
		\hline
		\# & Models & \multicolumn{3}{c|}{Goodness-of-fit} & \multicolumn{3}{c}{Predictive accuracy} \\
		&  & $R^2$ & RMSE & MAE & $R^2$ & RMSE & MAE \\
		\thickhline
1 & Generalized Linear Model & 0.507 & 0.265 & 0.206 & 0.470 & 0.268 & 0.211 \\
2 & Ridge Regression & 0.505 & 0.265 & 0.207 & 0.470 & 0.268 & 0.210 \\
3 & Lasso Regression & 0.487 & 0.270 & 0.209 & 0.459 & 0.271 & 0.211 \\
4 & Generalized Additive Model & 0.557 & 0.250 & 0.194 & 0.475 & 0.267 & 0.208 \\
5 & Multi Adaptive Regression Splines {[}degree=1{]} & 0.527 & 0.259 & 0.201 & 0.472 & 0.267 & 0.207 \\
6 & Multi Adaptive Regression Splines {[}degree=2{]} & 0.532 & 0.258 & 0.200 & 0.462 & 0.270 & 0.211 \\
7 & Multi Adaptive Regression Splines {[}degree=3{]} & 0.577 & 0.245 & 0.191 & 0.402 & 0.285 & 0.220 \\
8 & Multi Adaptive Regression Splines {[}degree=3; penalty=2{]} & 0.506 & 0.264 & 0.206 & 0.442 & 0.275 & 0.213 \\
9 & \textbf{Random Forest} & \textbf{0.886} & \textbf{0.127} & \textbf{0.098} & \textbf{0.437} & \textbf{0.276} & \textbf{0.217} \\
10 & Gradient Boosting Method & 0.887 & 0.126 & 0.100 & 0.365 & 0.293 & 0.229 \\
11 & Bayesian Additive Regression trees & 0.574 & 0.246 & 0.190 & 0.484 & 0.265 & 0.205 \\
12 & Null Model (Mean-only) & NA & 0.377 & 0.296 & NA & 0.369 & 0.292\\
\hline
	\end{tabular}
	\end{center}
	\end{adjustwidth}
\end{table}

\begin{table}[!ht]
\begin{adjustwidth}{0in}{0in} 
	\begin{center}
	\caption{\bf{Medium/Small Metropolitan Counties: Model performance comparison.}}
	\label{table:MSM_model_perf}
	\begin{tabular}{lp{7.2cm}|ccc|ccc}
	\hline
		\multicolumn{8}{c}{Medium/Small Metropolitan County Model} \\
		\hline
		\# & Models & \multicolumn{3}{c|}{Goodness-of-fit} & \multicolumn{3}{c}{Predictive accuracy} \\
		&  & $R^2$ & RMSE & MAE & $R^2$ & RMSE & MAE \\
		\thickhline
1 & Generalized Linear Model & 0.626 & 0.398 & 0.300 & 0.570 & 0.402 & 0.307 \\
2 & Ridge Regression & 0.626 & 0.398 & 0.300 & 0.570 & 0.402 & 0.308 \\
3 & Lasso Regression & 0.591 & 0.416 & 0.312 & 0.537 & 0.418 & 0.317 \\
4 & Generalized Additive Model & 0.779 & 0.305 & 0.233 & 0.645 & 0.364 & 0.274 \\
5 & Multi Adaptive Regression Splines {[}degree=1{]} & 0.750 & 0.325 & 0.249 & 0.655 & 0.359 & 0.272 \\
6 & Multi Adaptive Regression Splines {[}degree=2{]} & 0.760 & 0.319 & 0.246 & 0.627 & 0.371 & 0.280 \\
7 & Multi Adaptive Regression Splines {[}degree=3{]} & 0.790 & 0.297 & 0.230 & 0.587 & 0.391 & 0.287 \\
8 & Multi Adaptive Regression Splines {[}degree=3; penalty=2{]} & 0.724 & 0.340 & 0.264 & 0.617 & 0.379 & 0.286 \\
9 & \textbf{Random Forest} & \textbf{0.934} & \textbf{0.166} & \textbf{0.122} & \textbf{0.656} & \textbf{0.359} & \textbf{0.269} \\
10 & Gradient Boosting Method & 0.967 & 0.117 & 0.090 & 0.620 & 0.376 & 0.284 \\
11 & Bayesian Additive Regression trees & 0.804 & 0.287 & 0.218 & 0.667 & 0.354 & 0.266 \\
12 & Null Model (Mean-only) & NA & 0.650 & 0.456 & NA & 0.619 & 0.440 \\
\hline
	\end{tabular}
	\end{center}
	\end{adjustwidth}
\end{table}

From Table \ref{table:LCM_model_perf} that presents the performances  of the models developed for the large central metropolitan counties, we observe that random forest and gradient boosting method are the two most competitive algorithms that outperform all the other models in terms of goodness-of-fit. However, in terms of predictive accuracy, random forest outperforms the gradient boosting method.  Thus, we selected random forest model to capture and predict suicide mortality rate in the large central metropolitan counties.
Similar pattern can be found in Table \ref{table:MSM_model_perf} that exhibits the performances of the models developed for the medium/small metropolitan counties. Gradient boosting method tops the list in terms of goodness-of-fit followed by random forest, however random forest outperforms the gradient boosting method with regard to predictive accuracy. This phenomenon indicates gradient boosting method is overfitting the training data. Note that, BART model has a slightly higher predictive accuracy than random forest model, but it also demonstrates much higher loss in fitting training data to the model.
Therefore, we selected random forest as our final model to make further inferences of key socio-environmental impacts on suicide disparities in metropolitan areas.  

Compared to the ``null model" (a.k.a. ``mean-only" model), which is often used as a benchmark model in statistical analyses, the random forest algorithm offered an improvement of 66.3\% on in-sample RMSE and 66.9\% on in-sample MAE, while for the predictive accuracy it offered an improvement of 25.2\% on out-of-sample RMSE and 25.7\% on out-of-sample MAE for the large central metropolitan counties dataset. On the other hand, for the medium/small metropolitan counties dataset, the random forest algorithm provided an improvement of 74.5\% on in-sample RMSE and 73.2\% on in-sample MAE, while from the predictive accuracy perspective, it offered an improvement of 42\% on out-of-sample RMSE and 38.9\% on out-of-sample MAE. 



\subsubsection{Models' diagnostics} \label{section:model_diagnostics}
\begin{figure}[!h]
\centering
\includegraphics[width=0.6\textwidth]{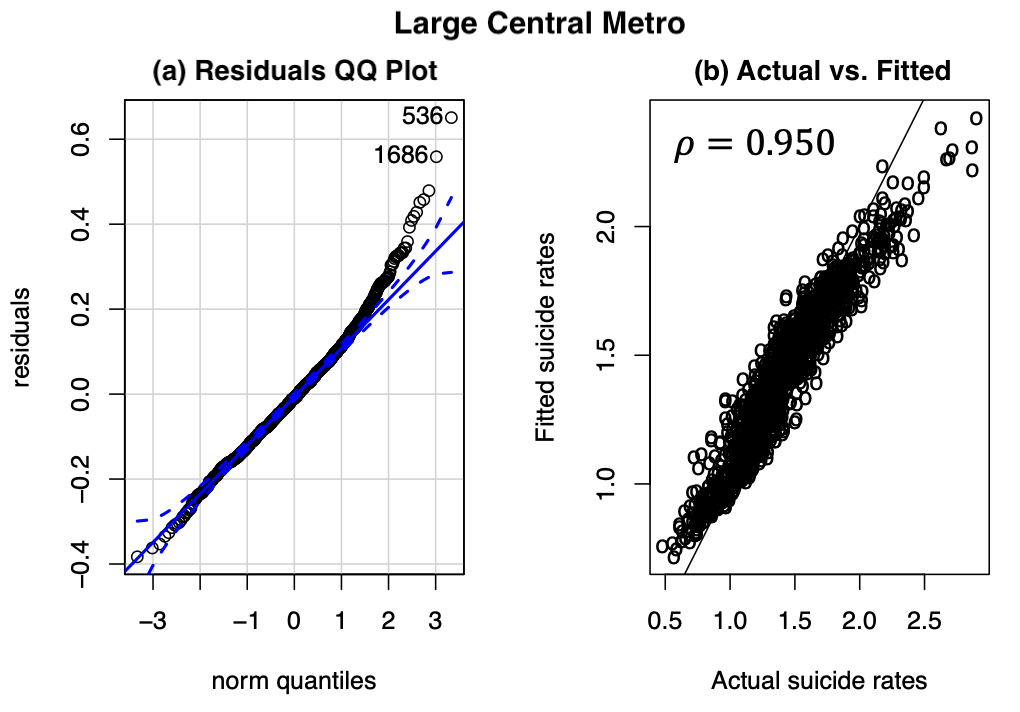}
\caption{{\bf Large Central Metropolitan Counties: Model diagnostics of final Random Forest model.}
(A) Residuals QQ plot (the blue dashed lines represent 95\% confidence intervals); (B) Predicted versus actual suicide counts, normalized per 100,000 of population.}
\label{fig6}
\end{figure}

\begin{figure}[!h]
\centering
\includegraphics[width=0.6\textwidth]{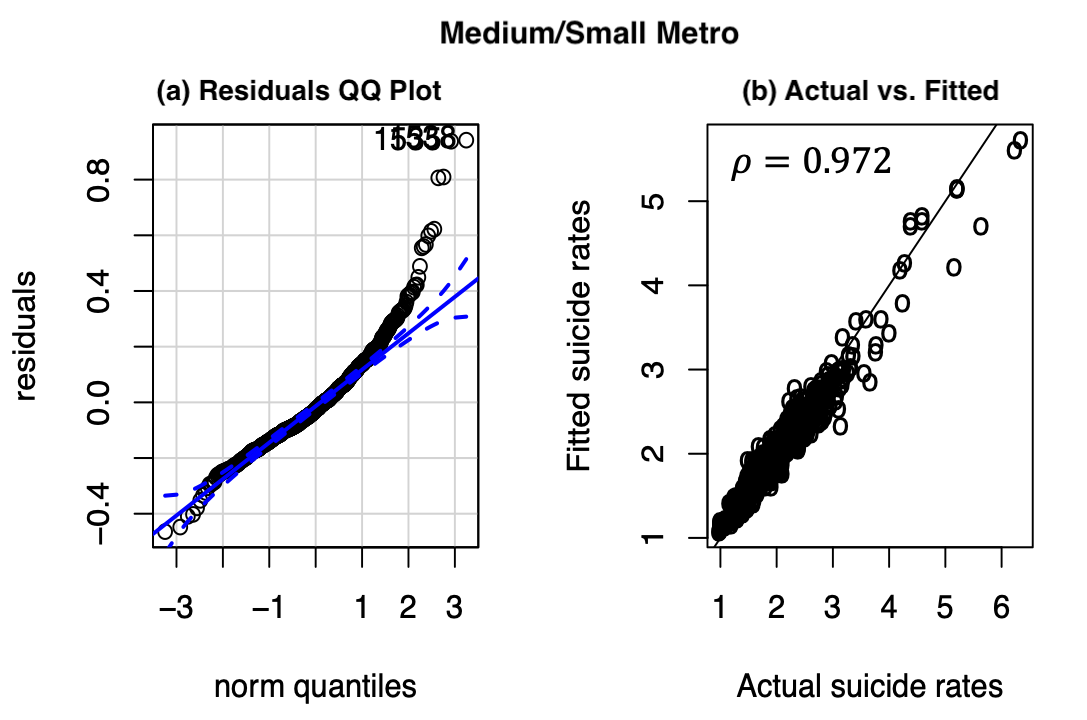}
\caption{{\bf Medium/Small Metropolitan Counties: Model diagnostics of final Random Forest model.}
(A) Residuals QQ plot (the blue dashed lines represent 95\% confidence intervals); (B) Predicted versus actual suicide counts, normalized per 100,000 of population.}
\label{fig7}
\end{figure}

To validate our finally selected random forest in capturing the suicide variations between both the large central and medium/small metropolitan counties, we analyzed the Q-Q plots of the model residuals as depicted in Figs.~\ref{fig6}(A) and~\ref{fig7}(A). A residual Q-Q plot is a graph that plots quantiles of the models' residuals versus quantiles of the standard normal distribution. From Figs.~\ref{fig6}(A) and~\ref{fig7}(A), we observe that the residuals mostly fall along the 45\degree line of the normal quantile plot, with slight deviations at the tails.  The deviated tails at the extremes indicate that there are unobserved heterogeneities, most likely associated with non socio-environmental factors (e.g., victim-level information on pre-existing clinical conditions, health behaviors, family issues, etc.) which could not be captured in our model. The higher $R^2$ values of the models---e.g., $R^2=0.886$ and $R^2=0.934$ for large central metro and medium/small metro counties models respectively, the higher values of Pearson correlation coefficients ($\rho$=0.950 in large central metro; $\rho$=0.972 in medium/small metro) between the actual and the fitted values, and the residuals Q-Q plots indicate that the selected random forest model can adequately capture the variation in the data and model the suicide mortality rates as a function of the various socio-environmental factors.

\subsection{Key predictors identification and ranking}
\label{section: variable_importance}
Variable importance is calculated based on ``variable inclusion proportion'', which is the fraction of times a given predictor was used in growing a regression tree (see more details in the description of random forest algorithm \ref{algo:RF}). In this paper, the variable importance ranking can be used to indicate the main influencing factors by their relevance of suicide mortality rates. For the sake of brevity, we selected top 15 variables in predicting suicide rates in the large central and medium/small metropolitan counties. Table \ref{table:summaryVar} exhibits those 15 variables and their sign of the correlation coefficients with the response variable.

\begin{table}[!ht]
\begin{adjustwidth}{0in}{0in} 
	\begin{center}
	\caption{\bf{Summary of top 15 variables in large central and medium/small areas.}}
	\label{table:summaryVar}
	\begin{tabular}{cl|cc|cc}
	\hline
		Variable & Description & \multicolumn{2}{c|}{Large Central Areas} & \multicolumn{2}{c}{Medium/Small Areas} \\ \cline{3-6} 
		& & Rank & Correlation & Rank & Correlation \\
		\thickhline
    AA & Percentage of Asian population. & 1 & Negative & 2 & Negative \\
    BA & Percent of Black population. & 12 & Mixed & 1 & Negative\\
    NH & Percent of non-Hispanic population. & 9 & Positive & 15 & Positive\\
    IA & Percent of  American Indian, Alaska native population. & 10 & Mixed & 5 & Negative\\
    NA &  Percent of Native Hawaiian, Pacific Islander population. & 13 & Positive & 6 & Negative\\
    Female &  Percent of female population. & 7 & Mixed & 4 & Negative \\
    Age\_1 &  Percent of young adults aged below 14 years old. & 6 & Mixed & 14 & Mixed \\
    Age\_2 &  Percent of adolescents aged 15-29 years old. & 4 & Positive & 8 & Negative\\
    Age\_6 &  Percent of elder people aged above 75 years old. & - & - & 7 & Mixed \\
    Education\_1 &  Percent of people with less than a high school degree. & 11 & Mixed & 11 & Positive \\
    Education\_2 &  Percent of people with a high school degree. & 8 & Mixed & 3 & Positive \\
    Education\_3 &  Percent of people with an associate degree. & 14 & Negative & 10 & Negative \\
    Unemployment &  Percent of unemployed workers in the total labor force. & - & - & 9 & Positive \\
    Income &  Median household income. & - & - & 13 & Mixed \\
    DX90 &  Number of days with temperature $\ge$ 90\si{\degree}F. & 2 & Positive & - & - \\
    DX70 &  Number of days with temperature $\ge$ 70\si{\degree}F. & 3 & Positive & - & - \\
    HDSD &  Seasonal heating degree days. & 5 & Mixed & - & - \\
    EMXP &  Extreme maximum daily precipitation total within month. & 15 & Mixed & - & - \\
    CDSD &  Seasonal cooling degree days. & - & - & 12 & Positive \\
\hline
	\end{tabular}
	\begin{flushleft} Note that, positive correlation denotes the relationship between predictor and response variable that changes in the same way (either increasing or decreasing), while negative correlation denotes this relationship changes in the opposite way. Otherwise, a mixed correlation indicates a combination of positive and negative relationship between predictor and response variable.
    \end{flushleft}
	\end{center}
	\end{adjustwidth}
\end{table}


\begin{figure}[!h]
	\begin{center}
	\includegraphics[width=0.5\textwidth]{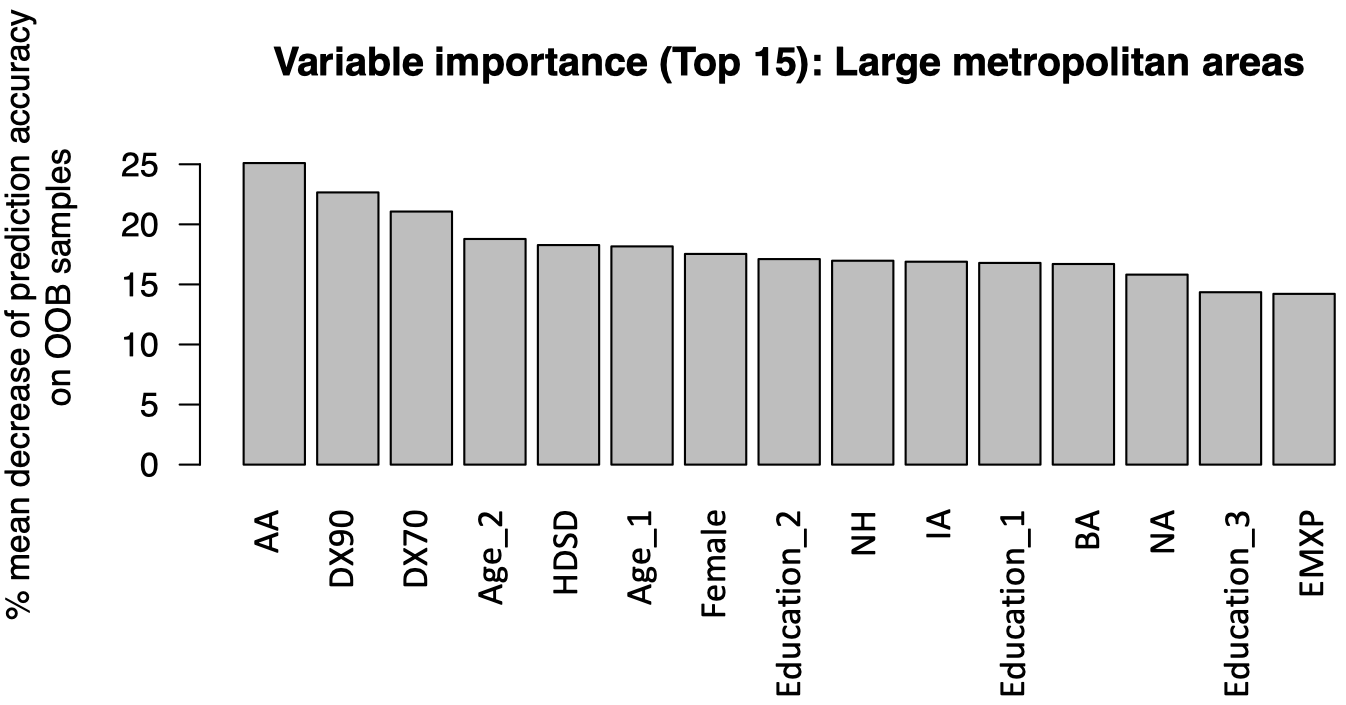}
	\caption{{\bf Large central metro: Variable importance ranking of top 15 predictors.} Variable descriptions are provided in Table \ref{table:summaryVar}.}
	\label{fig8}
	\end{center}
\end{figure}

\begin{figure}[!h]
	\begin{center}
	\includegraphics[width=0.5\textwidth]{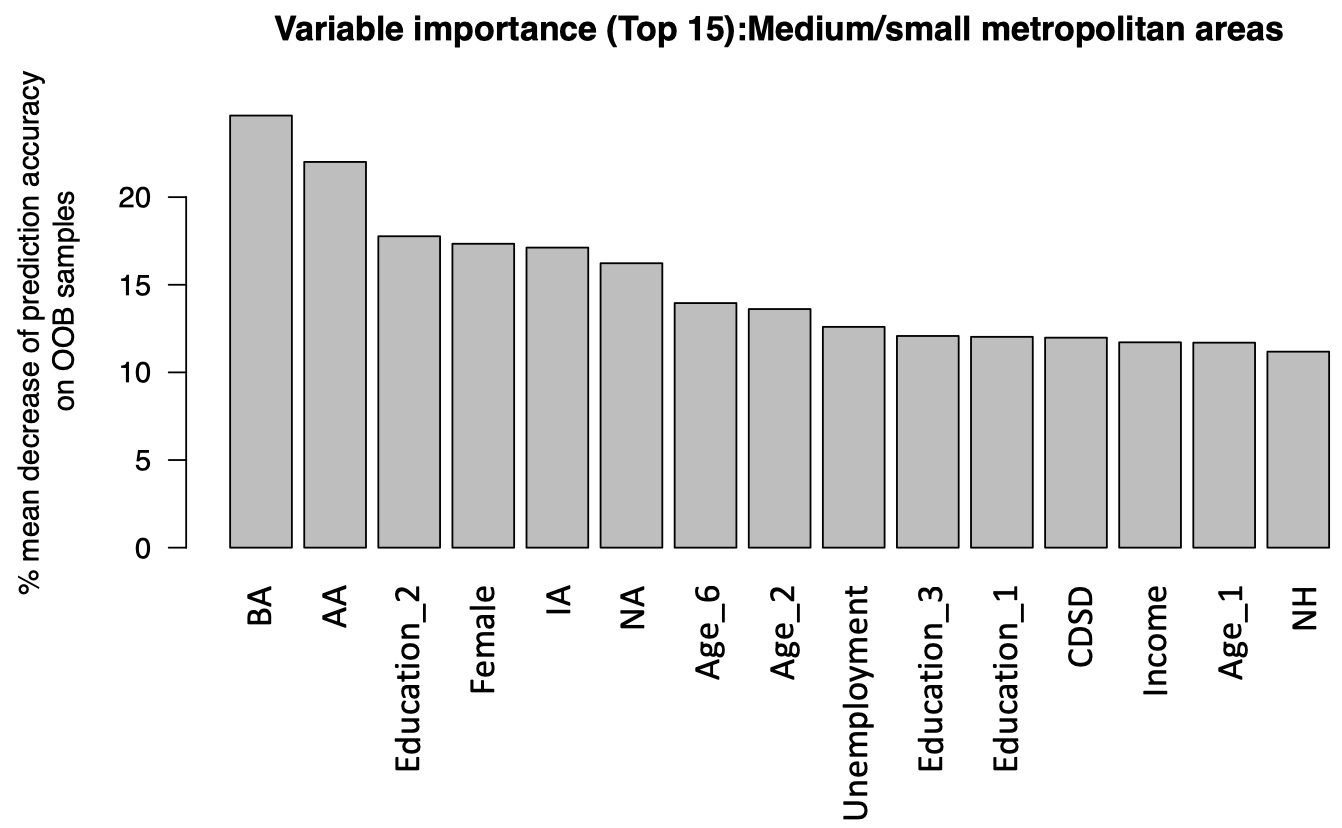}
	\caption{{\bf Medium/small metro: Variable importance ranking of top 15 predictors.} Variable descriptions are provided in Table \ref{table:summaryVar}.}
	\label{fig9}
	\end{center}
\end{figure}

As we observe from Figs. \ref{fig8} and \ref{fig9}, socio-demographic factors (race, gender, age, and education) have a different yet significant influence on the suicide mortality rate in both large central and medium/small metropolitan areas. Economic factors (i.e., unemployment rate and the median household income) have more impacts on suicide rates in the median/small metropolitan regions than in the large central metropolitan regions. In addition, suicide rates in the large central metros are found to be more sensitive to specific climatic variables (DX90---number of days higher than 90\degree F, DX70---number of days higher than 70\degree F, HDSD---season-to-date heating degree days, and EMXP---extreme maximum precipitation in a month); while suicides in the medium/small metros are more sensitive to season-to-date cooling degree days (CDSD). The rational behind our findings is explained in the following subsections.

\subsection{Model Inference: Comparative assessment across large central and medium/small metropolitan counties}
The disparities of suicide mortality rate across large central and medium/small metropolitan areas are examined based on the key factors identified in Table \ref{table:summaryVar}. The relative influences of those key socio-demographic, climatic and economic factors on the suicide mortality are illustrated using partial dependence plots (PDPs) (see the Equ. \ref{eq:PD}), where in each plot the y-axis represents the averaged suicide mortality rate influenced only by the predictor variable in the x-axis, considering all the other predictor variables to be constant \cite{greenwell2017pdp}. 

\subsubsection{Socio-demographic factors} 
A detailed insight on socio-demographic impacts on suicide rates across metropolitan areas is provided in the following. 

\noindent\textbf{1) Association of race and suicide mortality}\\
Specific racial groups such as Asian (AA), Black (BA), American Indians and Alaska natives (IA), native Hawaiian and Pacific Islander (NA) and Non-Hispanic (NH) are all ranked as top 15 predictors, but with different impacts on suicide rates across large central and medium/small metropolitan areas.  

\begin{figure}[!h]
	\begin{center}
	\includegraphics[width=1\textwidth]{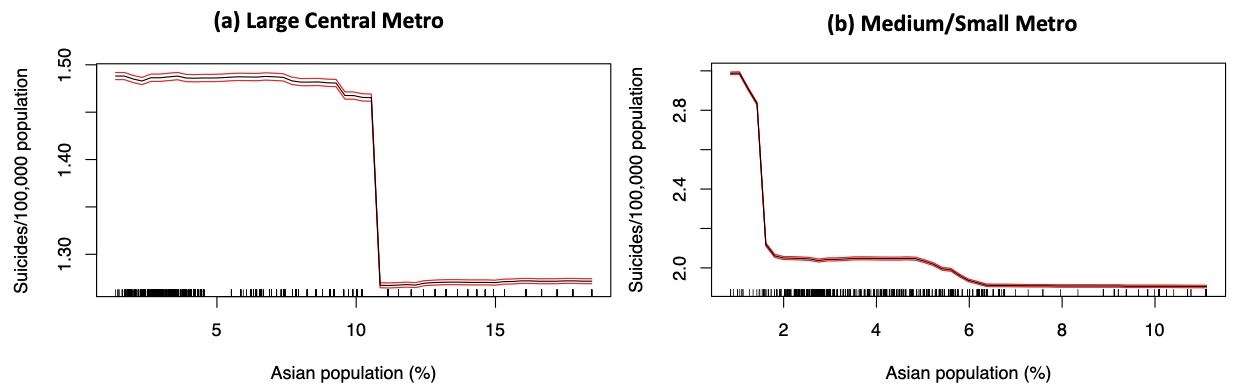}
	\caption{{\bf Suicide mortality rate in Asian population: (A) Large central metro; (B) Medium/small metro.} Rug lines on the $x$ axis indicate prevalence of data points; black curve is the average marginal effect of the predictor variable; red lines indicate the 95\% confidence intervals.}
	\label{fig10}
	\end{center}
\end{figure}

\begin{figure}[!h]
	\begin{center}
	\includegraphics[width=1\textwidth]{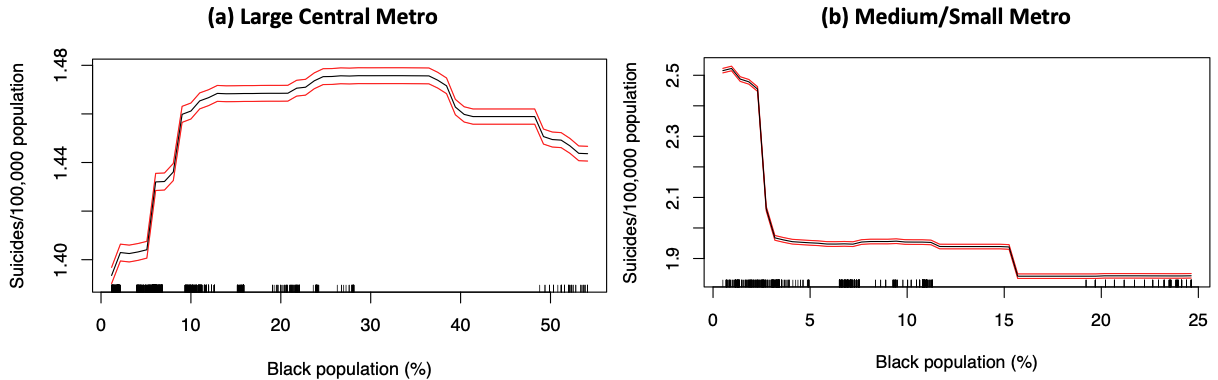}
	\caption{{\bf Suicide mortality rate in Black population: (A) Large central metro; (B) Medium/small metro.} Rug lines on the $x$ axis indicate prevalence of data points; black curve is the average marginal effect of the predictor variable; red lines indicate the 95\% confidence intervals.}
	\label{fig11}
	\end{center}
\end{figure}

The trend between the proportion of Asian population (AA) and suicide rate can be observed in Fig \ref{fig10}. This graph demonstrates that, as a growing AA, the averaged suicide rate first drops quickly and then stabilizes at a certain point. Specifically, suicide rate stabilizes at 1.28 per 100,000 population as the AA reaches above 11\% in the large central metros; while in the medium/small metros, suicide rates stabilizes at 1.9 per 100,000 population as the AA reaches above 6\%. This can infer that a community with higher Asian population (under certain threshold) could have a lower suicide rate. In general, Asian population is less likely to commit suicide. Previous study also indicated Asians were at low risk for suicide mortality compared to other racial groups such as White non-Hispanics and Black non-Hispanics \cite{duldulao2009correlates}. 
The association between the proportion of Black population (BA) and the suicide rates is demonstrated in Fig \ref{fig11}. In large metropolitan counties, we observed that suicide mortality rate increases as the BA grows, and declines as the BA exceeds 40\%. On the contrary, this relationship is different in the medium/small metropolitan areas, where a higher BA is related to a lower suicide rates---the average suicide rate goes down from 2.5 to 1.9 counts per 100,000 population of the county as the BA grows over 15\%. This opposite relation of suicide rates and Black population across different urbanized regions could be explained by the previous studies implying that Black population living in urban areas might feel more stressed and strained of the urban life due to unaccustomed social isolation or difficulty acculturating to middle-class suburban living \cite{willis2003uncovering,ivey2017suicide}. 

\begin{figure}[!h]
	\begin{center}
	\includegraphics[width=1\textwidth]{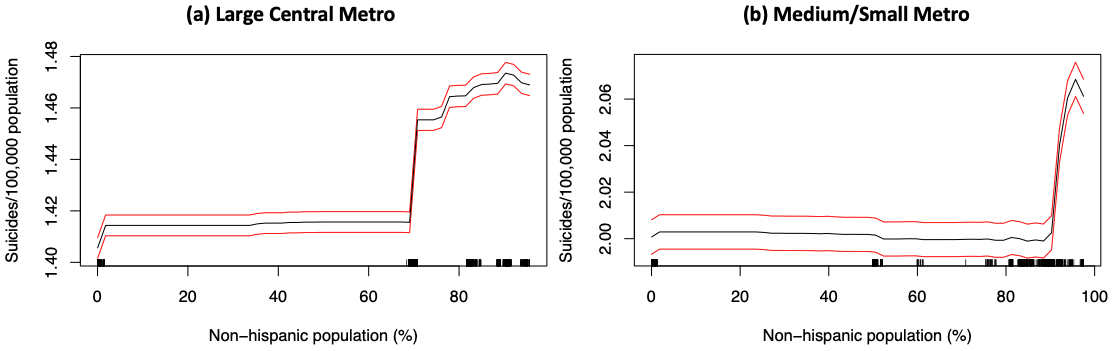}
	\caption{{\bf Suicide mortality rate in Non-Hispanic population: (A) Large central metro; (B) Medium/small metro.} Rug lines on the $x$ axis indicate prevalence of data points; black curve is the average marginal effect of the predictor variable; red lines indicate the 95\% confidence intervals.}
	\label{fig12}
	\end{center}
\end{figure}

\begin{figure}[!h]
	\begin{center}
	\includegraphics[width=1\textwidth]{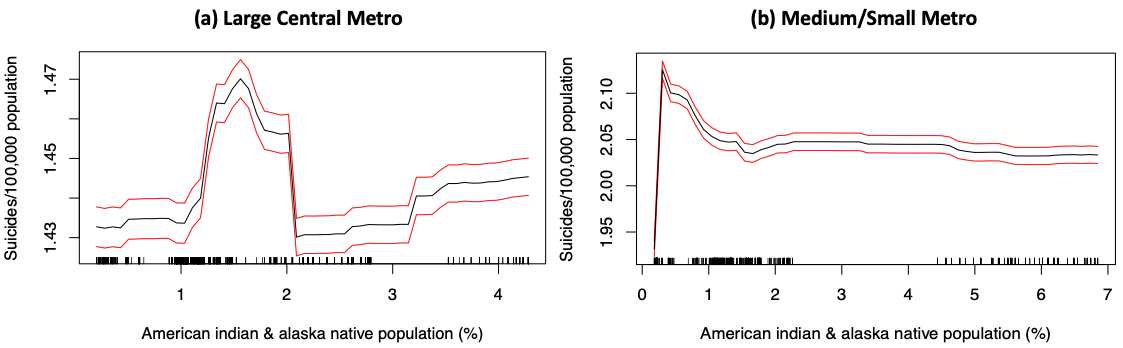}
	\caption{{\bf Suicide mortality rate in American Indian \& Alaska native population: (A) Large central metro; (B) Medium/small metro.} Rug lines on the $x$ axis indicate prevalence of data points; black curve is the average marginal effect of the predictor variable; red lines indicate the 95\% confidence intervals.}
	\label{fig13}
	\end{center}
\end{figure}

\begin{figure}[!h]
	\begin{center}
	\includegraphics[width=1\textwidth]{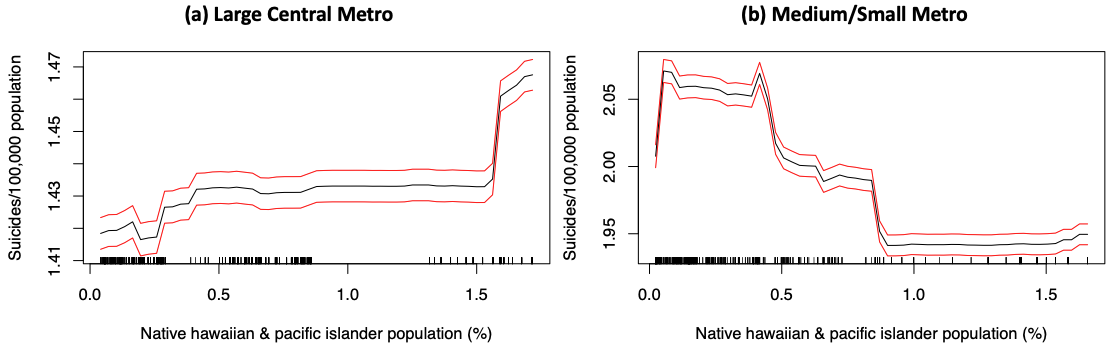}
	\caption{{\bf Suicide mortality rate in Native Hawaiian \& Pacific Islander population: (A) Large central metro; (B) Medium/small metro.} Rug lines on the $x$ axis indicate prevalence of data points; black curve is the average marginal effect of the predictor variable; red lines indicate the 95\% confidence intervals.}
	\label{fig14}
	\end{center}
\end{figure}

From Fig \ref{fig12}, we observe that higher proportion of Non-Hispanic population (NH) is also associated with increasing suicide mortality rate, and this relationship is consistent in both the large central and medium/small metropolitan counties. Note that, with a higher NH, the average suicide rate also increases. From the x-axis of Fig \ref{fig12}, the NH is a major group in the population and can account for the overall suicide rates, which is lined up with the existing research \cite{stone2018vital}.
Similarly, the relationship of suicide rates and the proportions of American Indian and Alaska natives (IA) can be observed from Fig \ref{fig13}. For the large central metros, the relationship depicts a step-function. More specifically, with IA ranging between 0.0---1.5\% and greater than 2\%, the suicide mortality rate shows an increasing trend, with an exception of a decreasing trend in the range of 1.5---2.0\. For the medium/small counties, we observe an increasing trend where the IA ranges between 0.0---1.0\%, but after that the trend is slightly decreasing.
Fig \ref{fig14} demonstrates the relationship of suicide rates and the proportions of Native Hawaiian or Pacific Islander (NA) population. More specifically, the suicide mortality rate shows an increasing trend with an increasing NA in the large central metropolitan counties, whereas, in contrast it shows a decreasing trend with an increasing NA in the medium/small metropolitan counties. 


\begin{figure}[!h]
	\begin{center}
	\includegraphics[width=1\textwidth]{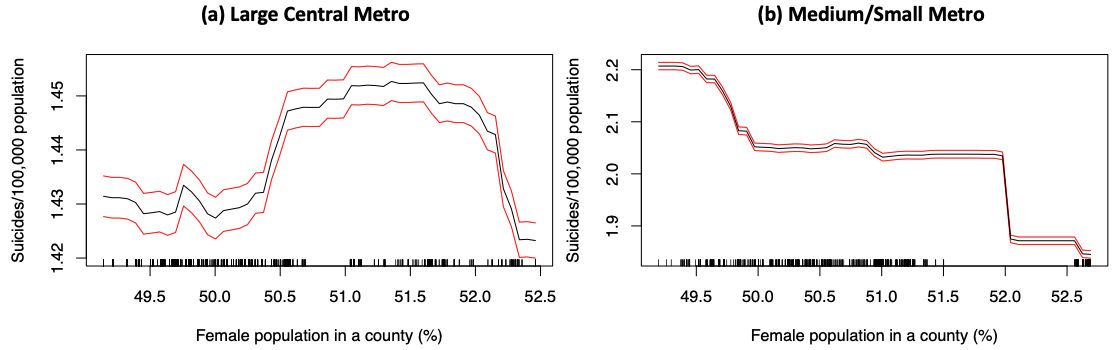}
	\caption{{\bf Suicide mortality rate among females: (A) Large central metro; (B) Medium/small metro.} Rug lines on the $x$ axis indicate prevalence of data points; black curve is the average marginal effect of the predictor variable; red lines indicate the 95\% confidence intervals.}
	\label{fig15}
	\end{center}
\end{figure}

\noindent\textbf{2) Association of gender and suicide mortality}\\
Gender plays a crucial role in understanding the variations in suicide mortality rates across the large central and medium/small metropolitan regions. From Fig \ref{fig15}, we found that in the large central metros, as the proportion of females increases, the suicide rate increases up until a certain point (around 1.45 per 100,000) and then it starts to drop sharply. In the medium/small metros, the suicide rate decreases monotonically as the growing of female populations. On average, counties having a higher proportion of females typically witness a lower suicide mortality rate. Previous studies stated that females have higher rates of suicidal ideations and attempts whereas males are more successful in completing a suicide, which is also known as the well-established concept of ``gender paradox in suicides" \cite{canetto1998gender,moore2018gender}.


\begin{figure}[!h]
	\begin{center}
	\includegraphics[width=1\textwidth]{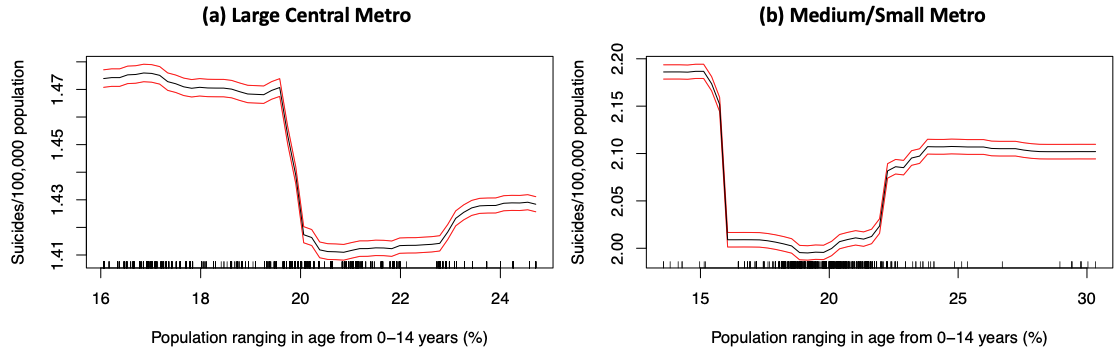}
	\caption{{\bf Suicide mortality rate among children ($0-14$ years age): (A) Large central metro; (B) Medium/small metro.} Rug lines on the $x$ axis indicate prevalence of data points; black curve is the average marginal effect of the predictor variable; red lines indicate the 95\% confidence intervals.}
	\label{fig16}
	\end{center}
\end{figure}

\begin{figure}[!h]
	\begin{center}
	\includegraphics[width=1\textwidth]{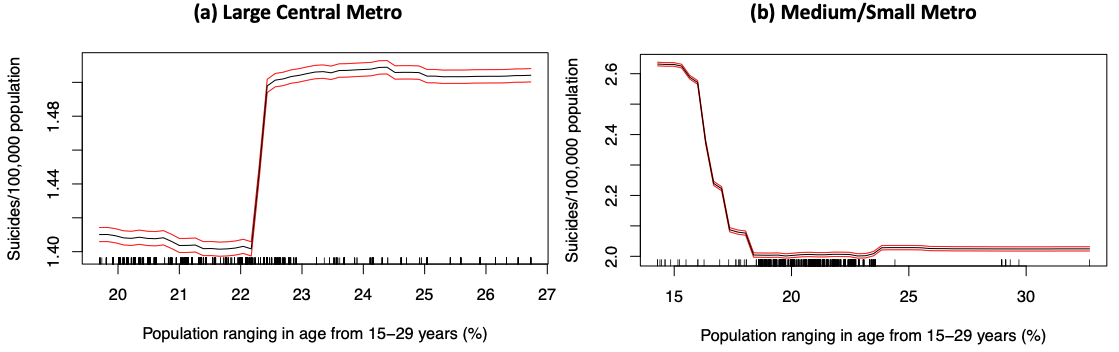}
	\caption{{\bf Suicide mortality rate among adolescents ($15-29$ years age): (A) Large central metro; (B) Medium/small metro.} Rug lines on the $x$ axis indicate prevalence of data points; black curve is the average marginal effect of the predictor variable; red lines indicate the 95\% confidence intervals.}
	\label{fig17}
	\end{center}
\end{figure}

\noindent\textbf{3) Association of age and suicide mortality}\\
Our studies also found that different age groups have certain impacts on suicide rates across the large central and medium/small metropolitan counties. In Fig \ref{fig16}, the suicide trend can be roughly represented in the form of step-function (decreasing first, reaches a plateau, and then increasing), with the increasing of the proportion of children and teenagers (aged under 14). And this trend can be observed in the both large central and medium/small metros. It is not surprising that prepubescent children are at risk of conducting suicidal behaviors, as previous studies suggested that by the age of eight or nine children have already formed a thorough understanding of suicide and do have intent to cause self-injury to possibly avoid their emotional pain such as break-ups \cite{tishler2007suicidal}. 
Fig \ref{fig17} relates to the proportion of adolescents aged between 15 to 29 years. In the large central metros, the suicide rate has a tipping point when adolescent population is around 22.2\% of the population. This indicates that suicide rate have a sharp increase from 1.4 to 1.5 per 100,000 population at the tipping point. Suicides among adolescents are growing in the last decades, and higher proportion of adolescents in the community could be linked to a higher suicide risks. Intriguingly, the suicide rate in the medium/small metros exhibits a downward trend with the increases as adolescent population grows. Note that, suicide rates eventually stabilize at 1.5 and 2.0 per 100,000 population for the large central and medium/small metros respectively, indicating the suicide disparities still need to be explained by other factors. 

\begin{figure}[!h]
	\begin{center}
	\includegraphics[width=0.5\textwidth]{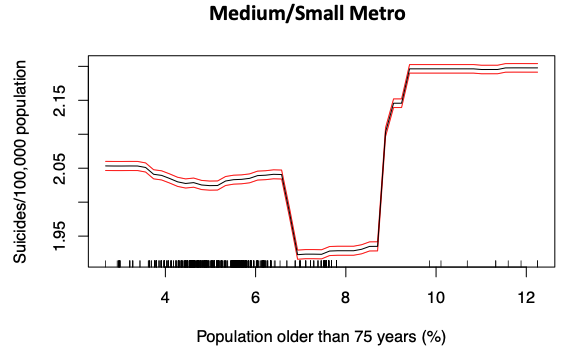}
	\caption{{\bf Suicide mortality rate among aged population ($\geq75$ years age) in Medium/small metro.} Rug lines on the $x$ axis indicate prevalence of data points; black curve is the average marginal effect of the predictor variable; red lines indicate the 95\% confidence intervals.}
	\label{fig18}
	\end{center}
\end{figure}

Our analysis also suggests that elderly people belonging to the age group of over 75 years and living in the medium/small metropolitan areas are vulnerable to committing suicides (see Fig \ref{fig18}). We observe that as the proportion of elderly population in a county increases beyond 7\%, the suicide mortality rate steadily increases. 
On the contrary, this factor does not appear to be significant (not ranked among the top 15 factors) for the suicide mortality rates in the large central metropolitan areas. Thus, the elderly population living in medium/small metropolitan settings has a higher risk of suicide, mostly due to the unavailability of sufficient mental health services or accessibility to the healthcare system \cite{eberhardt2004importance}.

\begin{figure}[!h]
	\begin{center}
	\includegraphics[width=1\textwidth]{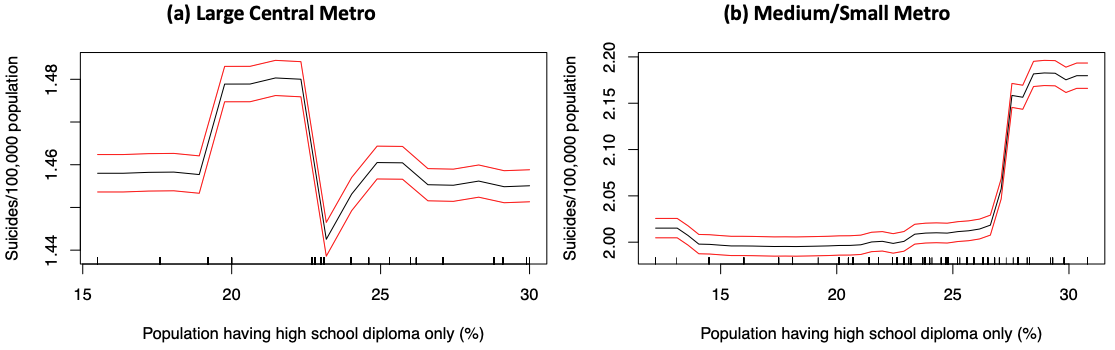}
	\caption{{\bf Suicide mortality rate among population having high school diploma only): (a) Large central metro; (b) Medium/small metro.} Rug lines on the $x$ axis indicate prevalence of data points; black curve is the average marginal effect of the predictor variable; red lines indicate the 95\% confidence intervals.}
	\label{fig19}
	\end{center}
\end{figure}

\begin{figure}[!h]
	\begin{center}
	\includegraphics[width=1\textwidth]{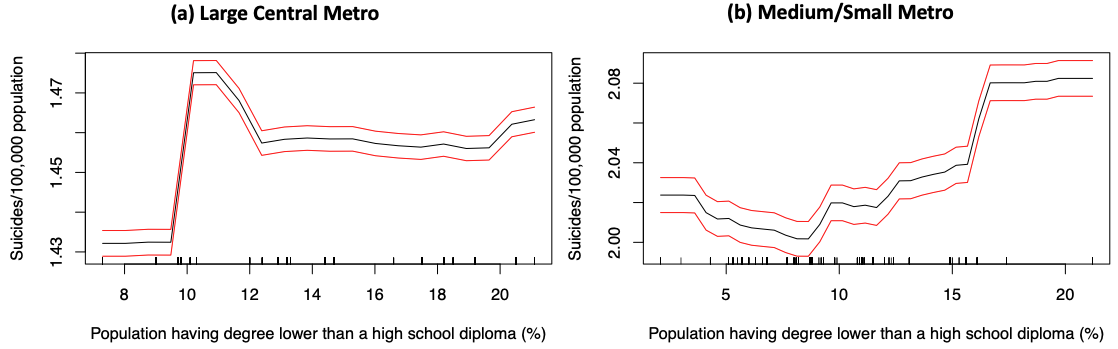}
	\caption{{\bf Suicide mortality rate among population having degree lower than a high school diploma): (a) Large central metro; (b) Medium/small metro.} Rug lines on the $x$ axis indicate prevalence of data points; black curve is the average marginal effect of the predictor variable; red lines indicate the 95\% confidence intervals.}
	\label{fig20}
	\end{center}
\end{figure}

\begin{figure}[!h]
	\begin{center}
	\includegraphics[width=1\textwidth]{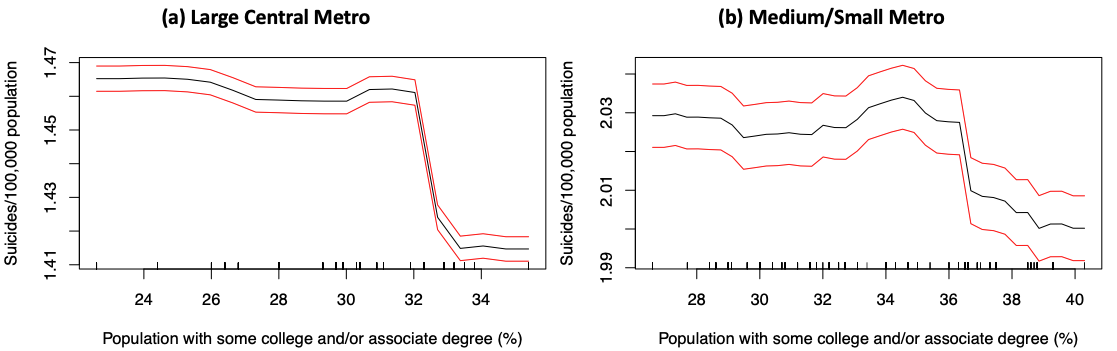}
	\caption{{\bf Suicide mortality rate among population having some college and/or associate degree): (a) Large central metro; (b) Medium/small metro.} Rug lines on the $x$ axis indicate prevalence of data points; black curve is the average marginal effect of the predictor variable; red lines indicate the 95\% confidence intervals.}
	\label{fig21}
	\end{center}
\end{figure}

\noindent\textbf{4) Association of education and suicide mortality} \\
The educational attainment of the population is an another key factor that can explain the suicide disparities across different urbanized regions. In the medium/small metropolitan counties, suicide rate shows an overall increasing trend with the growing proportion of people with a high school diploma only or lower (in Figs. \ref{fig19}(b) and \ref{fig20}(b)), but exhibits an decreasing trend with growing proportion of people having college associate degree (in Fig \ref{fig21}(b)); this indicates that people with a lower educational attainment living in the medium/small counties are more vulnerable to suicide risks. 
However in the large central counties, the partial effect of educational attainment on suicide rates is more fluctuating (in Figs. \ref{fig19}(a) and \ref{fig20}(a)). This can be mostly attributed to the fact that only few large central metropolitan counties contain higher percentages of population with lower educational attainment, thus the trend cannot be generalized. We also observe that the suicide rate has a steady downward trend as the percentage of people with college or associate degree increases in the large central counties (in Fig \ref{fig21}).

Based on those findings about educational attainment, our analysis demonstrates that people with low levels of education are more likely to be linked to higher suicide rates. This education gradient in the suicide mortality rate, in both the large central and medium/small metropolitan areas, can reflect the importance of education in changing the risk perception and health-related behaviors of a population which in turn could improve the overall mental health condition and emotional well-being of a community. This finding is consistent with a cross-national research report showing that the suicide rate is relatively high among group with only a high school degree, and relatively low among people having at least a college degree \cite{phillips2017differences}. To some extent, education is more than enriching knowledge, providing a platform/resource for individuals to improve their coping skills and maintain their physical, mental and social well-being.

\subsubsection{Economic factors}
\begin{figure}[!h]
	\begin{center}
	\includegraphics[width=1\textwidth]{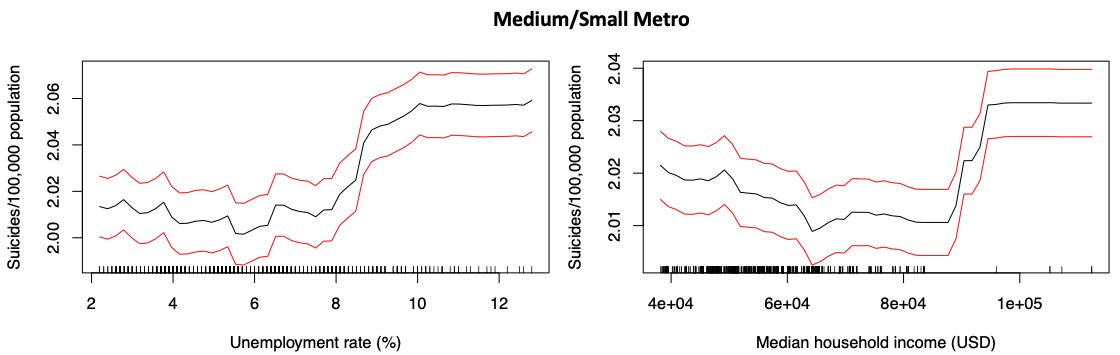}
	\caption{{\bf Association of economic factors with suicide mortality rates in medium/small metropolitan counties.} Rug lines on the $x$ axis indicate prevalence of data points; black curve is the average marginal effect of the predictor variable; red lines indicate the 95\% confidence intervals.}
	\label{fig24}
	\end{center}
\end{figure}

This study also reveals the association of economic factors and suicide mortality rates across the large central and medium/small metropolitan regions. We note that two economic factors (unemployment rate and median household income) that ranked as top 15 factors are of significant importance in influencing the suicide rates in the medium/small metropolitan areas. However, those two economic factors were not found to be important in the large central metropolitan areas. As presented in Fig \ref{fig24}, the suicide mortality rate has an increasing trend with the growing of unemployment rate in the medium/small metros. This finding is lined up with one previous research that examined an increase in the relative risk of suicide was linked to the unemployment status \cite{kposowa2001unemployment}. From Fig \ref{fig24}, we also found that the suicide mortality rate can be represented as a step-function of the median household income in the medium/small counties. The suicide mortality rate shows an decreasing trend as the median household income grows within the range from 40,000 to 80,000 USD annually. However, as the median household income increases above 80,000 USD annually, we observe an increasing trend in the suicide mortality rates. Note that, since only few observations fall in a range above 80,000 USD annually, we could consider that for the most cases, the relation between median household income and suicide rates are negative correlated if the median household income is below 80,000 USD annually. 

From the analysis, suicide rate in the medium/small metros is more sensitive to unemployment rate and median household income compared with suicide rate in the large central metros. Living in the medium/small metropolitan counties, people with less than average income or under unemployment may encounter more physical and mental stress, which could act as a trigger to underlying mental illness or chronic depression that can lead to committing suicide. While in the large central metros, there are more jobs opportunities.  
Moreover, some researchers found that people were being unemployment in a society may be considered as lack of social cohesion, which in turns is associated with the higher chance of committing suicide \cite{kawachi2000social, blakely2003unemployment}. 

\subsubsection{Climate factors}
In this study, we identified some climate factors that have a significant influence on the rate of suicides. Our findings can help understand how climate play a crucial role in affecting differences in suicide across metropolitan areas.  

\begin{figure}[!h]
	\begin{center}
	\includegraphics[width=1\textwidth]{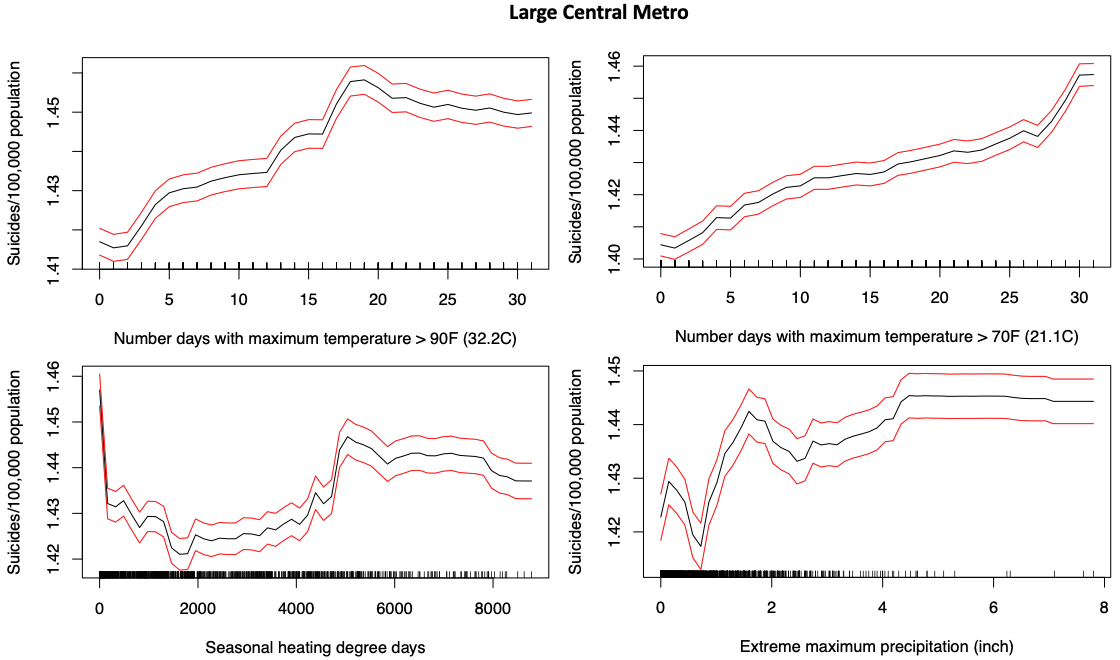}
	\caption{{\bf Association of climate with suicide mortality rate in large central metropolitan counties.} Rug lines on the $x$ axis indicate prevalence of data points; black curve is the average marginal effect of the predictor variable; red lines indicate the 95\% confidence intervals.}
	\label{fig22}
	\end{center}
\end{figure}

In large central metropolitan counties, the four climate factors that are ranked among the top 15 factors are DX90, DX70, HDSD and EMXP. Details of variable descriptions are provided in Table \ref{table:summaryVar}, and their relationships associated with the suicide mortality rate are exhibited in Fig \ref{fig22}. It can be observed from our analysis that the suicide mortality rate in large central metropolitan area has an increasing trend with higher extreme temperatures (i.e., DX90 and DX70). Our finding is consistent with prior research studies that claim a strong association between warmer temperatures and suicide rates \cite{dixon_association_2014, burke2018higher}. Suicide is known to have a strong correlation with persistent sadness and depression. One recent paper published in 2018 analyzed 600 million Twitter posts to connect mental health and temperature, and indicated that hotter months was linked to a higher chance in using depressive languages \cite{burke2018higher}. 
Additionally, our study also found that the seasonal heating degree days (HDSD) and the extreme daily maximum precipitation (EMXP) have positive associations with the suicide mortality rates. There is no existing research to examine the relationship between HDSD/EMXP and the suicide risks; however, one existing research have demonstrated that higher precipitation is linked with increasing mental health issues \cite{obradovich2018empirical}. It is well-recognized that mental health issues can contribute to suicidal behaviors; thus, it is not surprising to observe that EMXP has an indirect effect on higher rates of suicide mortality.

\begin{figure}[!h]
	\begin{center}
	\includegraphics[width=0.5\textwidth]{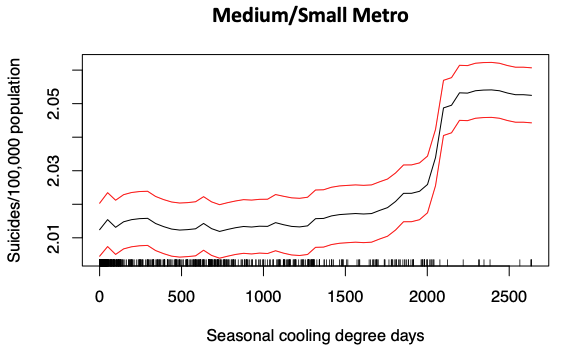}
	\caption{{\bf Association of climate with suicide mortality rate in medium/small metropolitan counties.} Rug lines on the $x$ axis indicate prevalence of data points; black curve is the average marginal effect of the predictor variable; red lines indicate the 95\% confidence intervals.}
	\label{fig23}
	\end{center}
\end{figure}

In contrast to the large central metropolitan areas, only one climate factor---CDSD (seasonal cooling degree days) appears in the list of the top key 15 factors, influencing the suicide mortality rates in the medium/small metropolitan areas. As can be observed from Fig \ref{fig23}, the suicide mortality rate has an upward trend with the increasing CDSD, clearly reflecting higher temperatures are associated with higher suicide mortality rates, similar to that of the large central metropolitan areas. Further investigation is needed to explain why the climatic conditions affect the suicide mortality rates of the geographic regions in a different way.

\section{Conclusion}
Evaluating the socio-environmental impacts on suicide mortality rate at the population level is critical to inform policy makers and healthcare providers in devising effective strategies that can help improve the mental health wellbeing as well as quality of life of residents across urban metropolitan counties. 
This paper investigated the association between socio-environmental factors (i.e., socio-economic, demographic, and climate) and suicide rates across the various geographical locations in the U.S. using a data-driven predictive approach, and assessed the key influencing socio-environmental factors that can best explain and predict the suicide disparities across the large central and medium/small metropolitan areas. 
Our research framework integrates novel statistical learning techniques that can not only identify the most influential features for prediction of the growth of suicide mortality rates, but also understand the relationships between these features and the suicide mortality rates. We implemented a library of parametric, semi-parametric and non-parametric statistical learning methods and found that the random forest algorithm best captures the underlying associations between the socio-environmental factors and suicide mortality rates. Random forest outperformed all the other predictive models that we implemented in this research in terms of in-sample goodness-of-fit and out-of-sample predictive accuracy. Our findings revealed that the interaction between suicide mortality rates and the socio-environmental factors varies significantly across the large central and the medium/small metropolitan areas in the U.S.

Our finding indicates that the population demographics plays a critical role in affecting the suicide rates across different metropolitan areas. Our study demonstrates that suicide rates in both the large central and the medium/small metros are sensitive to racial groups, proportion of females in the community, adolescents and adults aged below 29, and the groups of people with lower level of education attainment. In addition, the suicide rate in the medium/small metros is particularly sensitive to the elder people who are aged above 75 years. 

The influence of economic factors on suicide rates is demonstrated to be of more significance in the medium/small metropolitan counties than in large central metropolitan areas. Suicide rate in medium/small metros is particularly sensitive to the unemployment rate and median household income. We found that, with an increasing unemployment rate (from 2\% to 12\%), the suicide rate also increases (from 2.0 to 2.06 per 100,000 population); on the other hand, as the medium household income decreases from 80,000 USD to 40,000 USD, the suicide rates increases from 2.01 to 2.02 per 100,000 population in a county per month. Although this number seems small, it can account for a significant rise in the number of monthly suicides on average in the U.S., with a population of 310 million residing in the urban and rural areas. These economic factors also account for the disparity in suicide rates between the less and the more urbanized areas. The local government could make use of our findings to effectively subsidize public investments in less urbanized areas and/or provide government incentives to those population who are having financial difficulties.  

This study also illustrates that climatic variables are correlated with suicide risks. In the large central metros, the suicide mortality rate is more sensitive to higher temperature, seasonal heating degree days and extreme maximum precipitation; while in the medium/small metros, the suicide rate is more sensitive to seasonal cooling degree days. Our findings suggest that suicide rate increases with a higher temperature which is consistent with the previous studies \cite{dixon_association_2014, burke2018higher}. The weather variables may not account for direct motivation for people to commit suicide, but knowing the correlation between climate changes and suicidal variations is necessary to predict the future trend of suicide rates in the face of climate change. 

It is noteworthy that, although our data-driven framework was applied to model the suicide risks across the metropolitan areas, it is generalized enough to be able to conduct similar comparative assessments in other regions of the interests, provided relevant data is available. Although our study examined the socio-environmental impacts on suicide rates at the population level and can provide insights to make informed decisions at the community level, the findings may not be sufficient to make conclusions at the individual-level. Also, the proposed functions defining the relationships between suicide risks and socio-environmental factors could be beneficial for further investigation of suicide prediction and designing suicide prevention strategies. However, these relationships do not necessarily indicate causality. To reveal the causal relationship of suicide risks, the extensive longitudinal studies based on randomized control trials and other methodologies need to be conducted.

\section*{Acknowledgments}
The authors would like to acknowledge the 2019-20 SUNY Research Seed Grant Program for providing partial funding for this research. The authors would also like to thank McKenzie Worden and Andrew Kopanon---graduate students from the Department of Industrial and Systems Engineering at the University at Buffalo, The State University of New York for preliminary data collection for this research.

\section*{Appendix}
Table~S\ref{table:counties} lists all the counties along with their urbanization classification across the U.S., which were finally selected and considered in our analysis.

\begin{table}[!ht]
	\begin{center}
	\caption{Selected counties}
	\label{table:counties}
	\begin{tabular}{l l l}
		\hline
		County Name & State & Urbanization Level \\
		\hline
	    Mohave County & Arizona & Medium/Small Metro \\	
	    Maricopa County & Arizona & Large Central Metro \\	
                Pima County & Arizona & Medium/Small Metro\\
             Yavapai County & Arizona & Medium/Small Metro\\
              Adams County & Colorado & Large Central Metro \\	
          Arapahoe County & Colorado & Large Central Metro \\	
            Denver County & Colorado & Large Central Metro \\	
          Douglas County & Colorado & Large Central Metro \\	
            El Paso County & Colorado & Medium/Small Metro\\
         Jefferson County & Colorado & Large Central Metro \\	
              Weld County & Colorado & Medium/Small Metro\\	
                  Ada County & Idaho & Medium/Small Metro\\
                  Johnson County & Kansas & Large Central Metro \\	
            Sedgwick County & Kansas  & Medium/Small Metro\\	
         Jefferson County & Kentucky & Large Central Metro \\
          St. Charles County & Missouri & Large Central Metro \\	
         St. Louis County & Missouri & Large Central Metro \\	
      Jackson County & Missouri & Large Central Metro \\	
                Clark County & Nevada & Large Central Metro \\	
              Washoe County & Nevada & Medium/Small Metro\\	
 Hillsborough County & New Hampshire & Medium/Small Metro\\
      Bernalillo County & New Mexico & Medium/Small Metro\\	
    Oklahoma County & Oklahoma & Large Central Metro\\
             Tulsa County & Oklahoma & Medium/Small Metro\\	
      Clackamas County & Oregon & Large Central Metro \\	
          Multnomah County & Oregon & Large Central Metro \\	
          Washington County & Oregon & Large Central Metro \\	
          Davidson County & Tennessee & Large Central Metro \\	
          Shelby County & Tennessee & Large Central Metro \\
          Salt Lake County & Utah & Large Central Metro \\	
                  Utah County & Utah & Medium/Small Metro\\	
                 Weber County & Utah & Medium/Small Metro\\	
                 Clark County & Washington & Large Central Metro \\	
            King County & Washington & Large Central Metro \\	
        Pierce County & Washington & Large Central Metro \\	
      Snohomish County & Washington & Large Central Metro \\	
        Spokane County & Washington & Medium/Small Metro\\	
                          	\hline
	\end{tabular}
	\end{center}
\end{table}

\newpage


\end{document}